  \documentclass[aps,twocolumn,floatfix,,showpacs,superscriptaddress,footnoteinbib,prb]{revtex4-1}

\usepackage{amsmath, amsfonts, amssymb, mathrsfs, dsfont}
\usepackage{graphicx, color}
\usepackage{wasysym}
\usepackage{ulem} 
\usepackage{bm}
\usepackage{dcolumn}   
\usepackage[mathscr]{eucal}
\usepackage[dvipsnames]{xcolor}
\usepackage{caption}
\usepackage{subcaption}

\usepackage[colorlinks, breaklinks, 
            linkcolor=OrangeRed,
            citecolor=RoyalBlue,
            urlcolor=NavyBlue]{hyperref}
            
\usepackage[all]{hypcap} 

\newcommand{\be}{\begin{equation}}
\newcommand{\ee}{\end{equation}}
\newcommand{\bea}{\begin{eqnarray}}
\newcommand{\eea}{\end{eqnarray}}

\newcommand{\br}{{\bf r}}

\newcommand{\coloneq}{\mathrel{\mathop:}=}

\newcommand{\eq}[1]{\begin{align}#1\end{align}}

\newcommand{\Hsc}{\hat H_\mathrm{sc}}
\newcommand{\Hs}{\hat H_\mathrm{s}}
\newcommand{\Hh}{\hat H_\mathrm{H}}

\newcommand{\Hbdg}{{\hat H_\mathrm{BdG}}}
\newcommand{\Nbf}{N_\text{bf}} 
\newcommand{\calPld}{\mathcal {P_\text{ld}}}
\newcommand{\calPlg}{\mathcal {P_\text{lg}}}
 
\newcommand{\Philg}{\Phi_\text{lg}}
\newcommand{\Nc}{N_{\mathcal C}}
\newcommand{\NE}{N_{\text{E}}}
\newcommand{\Nnz}{N_\text{nz}}

\begin{document} 

\title{Self-consistent-field ensembles of disordered Hamiltonians: 
Efficient solver and application to superconducting films}

\author{Matthias Stosiek} 
\affiliation{ Institute of Theoretical Physics, University of Regensburg, D-93040 Germany}
\author{Bruno Lang}
\affiliation{ Institute of Applied Informatics, University of Wuppertal, D-42119 Germany}
\author{Ferdinand Evers}
\affiliation{ Institute of Theoretical Physics, University of Regensburg, D-93040 Germany}

\date{\today}

\begin{abstract}
  Our general interest is in self-consistent-field (scf) theories of disordered fermions. 
  They generate physically relevant sub-ensembles (``scf-ensembles'') 
  within a given Altland-Zirnbauer class.
  We are motivated to investigate such ensembles
  (i) by the possibility to discover new fixed points due to (long-range)
  interactions;
  (ii) by analytical scf-theories that rely on partial self-consistency approximations awaiting a numerical validation; 
  (iii) by the overall importance of scf-theories for the understanding of 
  complex interaction-mediated phenomena in terms of effective single-particle pictures.  
  
  In this paper we present an efficient, parallelized implementation solving 
  scf-problems with spatially local fields  by applying a kernel-polynomial approach.  
  Our first application is the Boguliubov-deGennes (BdG) theory of the attractive-$U$ Hubbard model in the 
  presence of on-site disorder; the sc-fields are the particle density $n(\br)$ and the 
  gap function $\Delta(\br)$. For this case, we reach system sizes unprecedented in earlier work.  
They allow us to study phenomena emerging at scales substantially larger than the lattice constant, 
such as the interplay of multifractality and interactions, or the formation of superconducting islands.
For example, we observe that the coherence length exhibits a non-monotonic behavior with increasing disorder strength already at moderate $U$. 
With respect to methodology our results are important because we establish that partial self-consistency 
("energy-only") schemes as  typically  employed in analytical approaches 
tend to miss qualitative physics such as island formation. 

\end{abstract}
\maketitle

\section{Introduction.}

The symmetry classification of disordered metals 
as it has been devised by Altland and Zirnbauer 
is nowadays considered to be complete.\cite{Zirnbauer1996ISB,Zirnbauer1997ISB,Heinzner2005ISB}
The classification is fundamental in the sense that 
all {\em generic} ensembles of random Hamiltonians have been covered. 
The classifying criterion is the presence or absence of one of the 
four elementary symmetries: time-reversal, 
spin-rotation, sublattice (chiral) and particle-hole (Boguliubov-deGennes-type). 

Based on an (incomplete) analogy to the conventional Landau-Ginzburg-Wilson 
theories of classical phase transitions, there was a wide-spread
misunderstanding with many researchers at the late 1980ies and early 1990ies
that a classification based on symmetry (and topology) alone  
would (more or less) determine the phase-diagrams and the associated critical points as well. 
In other words, the symmetry-classification was largely identified with a classification 
of universality classes, i.e. of all non-equivalent quantum field theories 
(low-energy action-functionals) that describe a disordered electron system. 
Therefore, it came as a surprise for the larger part of the community 
when models of disordered fermions had been found that 
formally belong to the same symmetry class but nevertheless exhibit different
phase diagrams.

It is perhaps fair to say that despite of the progress in the symmetry 
classification, we are still far from a systematic understanding of all 
universality classes and phase-diagrams that systems of disordered fermions
could exhibit. One could rephrase by saying that the generic ensembles of
random Hamiltonians covered in the ten-fold way possess physically relevant 
sub-ensembles that exhibit their own phase-diagrams and critical
fixed-points. The power-law random-banded matrices (PRBM) constitute a  well
studied example.\cite{Mirlin1996} 
It offers a laboratory for criticality that can be
addressed relatively easily with analytical and numerical techniques. 
\cite{Mirlin1996, Mirlin2000b} 

\subsection{General motivation for investigating scf-ensembles}

The appearance of criticality in the PRBM-ensemble is a synthetic property; 
it is imposed by putting long-range (power-law) correlations 
into the hopping amplitudes of a tight-binding Hamiltonian.
It therefore is interesting to explore properties of 
other ensembles that also exhibit long-range correlations in the Hamiltonian matrix
elements, but of a kind that is self-generated and in this sense "emergent". 
Plausible candidates for such Hamiltonians are 
effective single-particle systems that appear in self-consistent-field (scf-) theories of
interacting fermions. A prototypical example could be the Hartree-Hamiltonian of a 
disordered wire or film; it carries a long-range correlated on-site potential 
due to a weakly screened Coulomb-interaction. 
  
Quite generally, we have in mind fermionic Hamiltonians 
\begin{equation}
\label{e1}
\Hsc = \frac{1}{2} \sum_{xy} \left[ h_{xy}[n,\Delta] \ c^\dagger_{x}c_y +
\Delta_{xy}[n,h]\ c^\dagger_{x}c^\dagger_{y} + \text{h.c.} \right]; 
\end{equation}
the matrix $h_{xy}[n,\Delta]$ is a functional of the density matrix $n$ and the pairing fields $\Delta$.
The self-consistency condition inherent to generic mean-field theories
requires that the fields $n$ and $\Delta$ are expectation values of operators 
$\hat n, \hat \Delta$ to be calculated employing - amongst other ingredients, 
such as density matrices or exchange-correlation kernels -- also $\Hsc$. 
Thus, scf-conditions are implied,  
\begin{equation}
n=\langle \hat n \rangle_{h,\Delta},\quad \Delta = \langle \hat \Delta \rangle_{n,h} 
\label{e2}
\end{equation}
that $h,\Delta$ and $n$ obey. 

A microscopic randomness will enter $\Hsc$, e.g., via $h$ incorporating random
on-site energies or hopping amplitudes.
The set of random Hamiltonians introduced thereby follows the conventional
symmetry classification. However, only the subset of all
members of a given symmetry class that happens to comply 
with Eq. \eqref{e2} forms the {\em scf-ensemble}.

We believe that scf-ensembles, their physical and 
mathematical properties constitute a fundamental research topic
that may not yet have received the amount of attention 
it deserves. Our belief bases on two observations: 
(i) The elements of scf-ensembles certainly tend to exhibit non-trivial
correlations in their matrix elements $h_{xy}$ and $\Delta_{xy}$. 
If correlations happen to be strong enough, e.g. sufficiently 
long ranged, then new phases with novel critical behavior can be 
expected to emerge. 
(ii) Mean-field theories are important because they provide a 
tractable reference point for a perturbative analysis of 
interaction effects. Thus, they are a generic encounter in all 
theories of disordered fermions that try to incorporate interactions.
To reveal, in particular, the impact of quantum fluctuations a 
thorough understanding of the mean-field reference point would certainly 
seem helpful. 

We give examples for occurrences of scf-ensembles:
\begin{description}
\item[Hartree theory (H)] The obvious example to define ensembles of
self-consistent Hamiltonians would be the Hartree-theory. In this case 
$\Delta{=}0$ and the field $n$ in Eqs. \eqref{e1} and \eqref{e2} 
should be identified with the particle density
$n(\br)$.
 \item[Hartree-Fock theory (HF)] $\Delta{=}0$; $n$ resembles the density matrix $n(\br,\br')$
and $h$ the Fock-operator. 
\item[Density-functional theory (DFT)] In the orthodox flavor 
$\Delta{=}0$, $n$ represents the particle density $n(\br)$ and $h$ becomes the
Kohn-Sham-Hamiltonian. Roughly speaking, 
DFT differs from HF due to the presence of
correlations in $h[n]$.
\item[Boguliubov-deGennes-Hamiltonian (BdG)] The basic 
Hamiltonian is given in Eqs. \eqref{e1} and \eqref{e2}.  
\end{description}

Our short list is far from exhaustive and further examples could be given.
For instance, we recall that many spin-systems have faithful
representations in terms of fermionic
network-models that also could be dressed with self-consistency requirements,
like self-consistent fluxes. 

\subparagraph*{Remarks:} 
(i) The investigation of scf-ensembles  is 
a very challenging endeavor. The difficulty is that each disorder configuration
requires to find its own self-consistent fields $h$ and $\Delta$. 
The solution of the scf-cycle is very difficult to do with analytical techniques. 
But also numerically it is demanding already at moderate system sizes of a few
thousand sites. 
Consequently, the number of studies including full self-consistency 
appears to be limited.
In Tab. \ref{t1} we list contributions most relevant to us.

(ii) A more general perspective can be developed 
that operates with self-consistency constraints on the Green's function 
$G(E)$ rather than on the elements of the Hamiltonian. 
The generalization becomes non-trivial when the 
self-energy $\Sigma(E)$ picks up an energy dependence. 
As a prototypical example we mention the GW-theory. 
It constitutes an electronic-structure method
that builds upon the Hedin-equations 
approximating them by ignoring vertex 
corrections.\cite{hedin1965}
In its full flavor the theory features a 
Green's function that satisfies a self-consistent set of 
equations defined by a truncated diagrammatic expansion. 
\cite{bechstedt,vanSetten2013}

\begin{table}[bp]
{\small
\begin{tabular}{|c|c|c|c|c|} \hline\hline
dim & mean-field & observables & parameters & Ref.\\
\hline\hline 
IQHE & HF & Thouless numbers, IP & $L\approx 10$ & \cite{yang95} \\ 
\hline
&TDHF& "Kubo conductivity" & &\cite{backhaus99} \\ \hline\hline
3D & HF & DoS, IP & $L=10$ & 
\cite{epperlein97} \\ \hline
3D&HF& DoS, IP& $L=24$ & \cite{yang95} \\ \hline
3D  & HF & mf-dim $D_2$ & $L=10$ &  \cite{amini14} \\ \hline
3D  & HF & DOS, mf-dim $D_2$ & $L=18$ &  \cite{Lee2018} \\ 
  \hline
3D & BdG & $\Delta(\br)$, LDoS & $L=50 (3D)$ & 
\cite{peeters2010}\\
& s-wave & &  & \\\hline
3D & DFT &  mf-spectrum $f(\alpha)$ & $L=22$ & 
\cite{carnio2019}\\ \hline
3D & LDA & $\nu$ with  & factor of&  \cite{harashima12,harashima14} \\ 
& &KS-states & two in $L$  &   \\ \hline\hline
2D & BdG & DoS, $P(\Delta)$ & $L=24$ & \cite{ghosal1998,ghosal01}\\ 
& s-wave & &&    \\ \hline
2D & BdG & DoS, $P(\Delta)$ & $L=38$ & \cite{Potirniche2014}\\
& s-wave & LDoS & & \\\hline
2D & BdG & DoS, $\Delta(\br)$ & $L=36$ & \cite{ghosal04,ghosal17, ghosal17b}\\ 
& s-, d-wave & &&    \\ \hline
2D & BdG & $P(\Delta)$, $\Delta(\br)$ & $L=25$ & \cite{castellani12, castellani13, castellani14, castellani15}\\ 
& s-wave &$\Phi_{AB}(\mathbf{q})$ &&    \\ \hline
2D & BdG & $\Delta(\br)$ & $L=12$ & \cite{dubi07}\\ 
& s-wave & &&    \\ \hline\hline
\end{tabular}
\caption{\label{t1} Studies of mean-field Hamiltonians 
in the literature that have been 
performed with full self-consistency. 
Abbreviations: Integer quantum Hall effect (IQHE); Hartree-Fock (HF); Time-Dependant Hartree-Fock; density of states (DoS); 
inverse participation ratio (IP);
multifractal (mf); 
local density approximation (LDA) for density functional theory in Kohn-Sham
(KS) formulation; 
localization length exponent: $\nu$; 
linear system dimension: $L$;
Boguliubov-deGennes-type pairing terms (BdG); 
distribution function of local SC-gaps:  $P(\Delta)$;
Correlation Function of various observables $A$ and $B$: $\Phi_{AB}(\mathbf{q})$.
}}
\end{table}

(iii) A potential classification scheme of scf-ensembles will involve concepts 
very different from the one designed by Altland and Zirnbauer (AZ).
To see this, we recall that AZ distinguish ten classes 
according to presence or absence of discrete symmetries.
In contrast, the scf-requirement as formulated in \eqref{e2} 
invokes parameter-bound kernels. Hence, 
{\it a priori} the number of scf-ensembles is not limited and 
an impression might arise according to which the scf-ensembles carry 
a degree of arbitrariness and therefore are less fundamental.
To address this reservation against the basic concept, 
we recall that there is a very special set of scf-theories which is standing out; 
these scf-theories share with the parent field theory they derive from 
the basic symmetries and in this sense are conserving.\cite{kadanoff} 
Therefore, a classification of {\it conserving scf-ensembles} goes together with 
the basic program of condensed-matter theory, which is to identify and understand 
the fixed-point theories that are possible within a given AZ symmetry class. 

\subsection{Motivation for numerics and challenges} 
 The numerical challenge that the scf-ensembles pose as compared to
 simulations of noninteracting fermions is that for each 
sample the scf-equation Eq.\eqref{e2} has to be solved in an iterative
fashion. Since the ensemble average requires solving  
hundreds of samples, typically, the computational cost
for such studies is extensive.
Presumably, this is the main reason why numerical studies of 
scf-ensembles have been performed infrequently in the past, 
despite of their obvious fundamental relevance. 

Thus motivated, we here present an implementation of the scf-problem 
that allows to achieve relevant system sizes at an affordable numerical cost. 
The interplay of disorder induced quantum-interference and mean-field interactions can be 
studied on length scales that exceed the lattice constant by two orders of magnitude.

\subparagraph*{Reduction of scaling - KPM:}
The computationally demanding step limiting the code-performance 
is  the calculation of the scf-fields, $n$, $\Delta$ and $h$, 
that need to be evaluated in every iteration cycle of the self-consistency process. 
In the case of Hartree-Fock-theory, for instance,  this implies the
reconstruction of the density-matrix from a given Fock-operator. 
In straight-forward implementations the Hamiltonian is diagonalized 
in each iteration cycle to feed eigenvalues and 
eigenvectors into the rhs of \eqref{e2}; the cost is 
${\cal O}(\Nbf^3)$ operations, 
where $\Nbf$ is the dimension of the single-particle Hilbert space.
\footnote{Once the scf-field was found an update of $h$ has to be
computed. This computation is efficiently dealt with by 
employing the fast Fourier-transformation (FFT) and therefore not critical. 
With FFT an operation that formally is ${\cal O}(\Nbf^2)$ can be downgraded to
${\cal O}(\Nbf\ln \Nbf)$.}

Consider the Hartree-approximation: 
the matrix diagonalization appears, because the trace 
\begin{equation}
\label{e4}
n(\br) = \mathrm{tr} \left[ f(\mu- \Hh) \hat n(\br) \right] 
\end{equation}
contains 
the Fermi-Dirac function, $f(\mu-\Hh)$, 
of a matrix valued argument that conventionally 
is evaluated in the basis of eigenstates of 
$\Hh$.

What many suggestions for ${\cal O}(\Nbf^x)$-solvers, $x<3$, of the
self-consistency problem have in common is that they 
employ an alternative approach for trace-computation
that avoids a diagonalization of $\Hh$ and therefore can be
more efficient, in principle, than ${\cal O}(\Nbf^3)$.
One of the well established options is the  kernel-polynomial method 
(KPM). \cite{weisse06} 
The conceptual idea behind this approach is to 
expand $f(x)$ into a rapidly converging series of 
Chebyshev polynomials, $T_l(x)$, that are obtained recursively: 
\begin{eqnarray}
\label{e4}
 n(\br) &=& \sum_{l=0}^{N_{\cal C}} a_l \text{tr } T_l(\Hs) \hat
n(\br)\nonumber\\
\label{e5}
        &=& \sum_{l=0}^{N_{\cal C}} a_l \sum_{b=1}^{\Nbf}  
            \langle b| T_l(\Hs) \hat n(\br)| b\rangle ,         
\end{eqnarray}
where $\Hs$ denotes an appropriately 
scaled Hamiltonian $\Hh$ and $|b\rangle$ is a suitable 
basis in which $\Hs$ 
is sparse.
($\Nc$: order of the Chebyshev expansion; $a_r$: known expansion coefficients;
$\Nbf$: number of basis functions). 
As is seen from \eqref{e5}, 
 the evaluation of the trace is of order ${\mathcal O}(\Nc \Nbf^2\Nnz)$. 
$\Nnz$ denotes the number of non-zero entries of $\Hs$ per row. 
For a dense matrix we have $\Nnz{=}\Nbf$, while for a very sparse matrix 
$\Nnz \approx \Nbf^{0}$. For example, 
 the BdG case, we have $\Nnz = 2d + 2$ , with $2d$ denoting the number of nearest neighbors 
on a cubic lattice in $d$ dimensions. 


\subparagraph*{Signatures of  implementation:}
We  have implemented a matrix-free KPM-solver of the 
self-consistency problem \eqref{e1}, \eqref{e2} 
for the situation where the self-consistent 
fields are local in real space $n(\br)$ and $\Delta(\br)$,
as is the case for the Hartree approximation and the Boguliubov 
theory of s-wave superconductors.
It operates at zero and non-zero temperature and is optimized, 
for accelerated convergence for averages over the phase-space 
of disordered scf-ensembles. 

The KPM-aspect of our implementation is similar to other variants 
described in earlier work. They have been proven 
useful in applications of the BdG-equation for nanostructures 
with one or very few impurities, but have not been applied to disordered 
samples. Implementation differences are in details: 
\textcite{peeters2010,nagai2012,nagai2013} also use KPM to perform 
traces. In addition, \textcite{peeters2010} also have employed a matrix-free implementation. 
While these authors expand the Green's function employing 
the Lorentz kernel, we expand the spectral function 
where the Jackson Kernel typically has better convergence 
properties \cite{weisse06}.

\subsection{Application to dirty superconductors}

Motivated by experiments on the superconductor-insulator
transition, e.g. Ref. \textcite{baturina07} and \textcite{sacepe08},  
the attractive$-U$ Hubbard model with on-site disorder 
has been studied recently in several 
computational\cite{dubi07, trivedi11, castellani12, castellani13, castellani14, Potirniche2014, castellani15, castellani15b, scalettar16}
and analytical works.\cite{Feigelman2007ISB, feigelman10,  feigelman15}

Important insights have been gained within the framework of BdG-theory. 
The most striking findings include  
(i) the  granularity of the pairing amplitude ("islands") emergent
on the scale of the coherence length even for short-range disorder\cite{ghosal01};
(ii)  the parametric decoupling of the spectral gap from the mean pairing amplitude 
at large disorder: while the first remains relatively large the second decays to zero. \cite{ghosal01}
(iii) A parameter regime was predicted 
where the typical size of pairing amplitude is increased as compared to the clean limit, 
so disorder has a pronounced tendency to {\em enhance} superconductivity. 
The mechanism was explored in 3D near the Anderson transition\cite{Feigelman2007ISB,feigelman10} 
but also in 2D samples with short and long-range 
interactions\cite{Burmistrov2012ISB,Burmistrov2013,Burmistrov2015ISB}. 
Several predictions are broadly consistent with numerical results obtained 
on a honeycomb lattice \cite{Potirniche2014}.  
(iv) At large interactions the coherence length was reported to exhibit a non-monotonous behavior
with increasing disorder strength.\cite{castellani15}

Despite the progress the current situation is not fully satisfying: 
On the one hand side, computational studies of BdG-Hamiltonians 
have been limited to system sizes
$L$ that do not allow to study the most interesting regime of length scales 
where the coherence length exceeds the lattice spacing: $\xi\gg a$. 
While analytical approaches, on the other hand, operate in this regime, 
they rely on partial self-consistency in order to become tractable.

Motivated by this observation, we investigate as a first application of 
our technology the BdG-problem of disordered superconductors
focussing on $s$-wave pairing in thin films at $T{=}0$. 
The full parameter plane of disorder, $W$, and interaction, $U$, 
is considered in which we study the distribution function and autocorrelations 
of the local gap function, $\Delta(\br)$, as our main observable.
Our computational machinery allows us to cover the full parameter space 
from the extreme regimes, which have been addressed computationally 
before, to the analytically tractable weak coupling limit. 
In this effort we observe the formation of islands in large regions of the parameter space, 
for the first time on mesoscopic scales considerably exceeding the lattice constant.
Regimes are included with parameters relatively close to the one 
where strong inhomogeneity has been observed experimentally.\cite{sacepe2011}.

Our observation might indicate that islands  play a crucial role for the stability of the 
superconducting phase in actual experiments.
Namely, islands imply localized Cooper pairs 
and therefore a diminishing of the phase-stiffness. In other words, 
islands go together with enhanced phase-fluctuations that 
destabilise long-range superconducting order. This connection  
between island-formation and stability has been emphasized before.\cite{ghosal01, dubi07} 

Calculating the autocorrelation function of the spectral gap, 
$|\Delta({\bf q})|^2$ we can extract a characteristic inverse length 
scale $\xi^{-1}(W,U)$ with the physical meaning of a correlation length. 
We study $\xi$ within the full phase diagram. 
Interestingly, concomitantly with island formation we find an enhanced 
BdG-coherence length. A similar observation if only at very large interaction strength, 
$U{=}5$, has been made by \textcite{castellani15}.  
To what extent the enhancement of $\xi$ is 
an artefact of mean-field theory that is removed when adding phase fluctuations
remains to be seen. 

Finally, like earlier authors\cite{ghosal01} we also pay a special attention to 
the sensitivity of the behavior of computational observables 
to approximations made in the self-consistency procedure. 
We find that the island formation when observed in moderate parameter regions 
is a characteristic hallmark of full self-consistency; it escapes 
partial (``energy-only'') self-consistent schemes. We conclude that 
the renormalization of wavefunctions associated with full self-consistency will 
probably be an important ingredient of a qualitative 
theory of the superconductor-insulator transition.

\section{Model and Implementation}

\subsection{BdG-treatment of Hubbard model}  
 
We consider the attractive$-U$ Hubbard model \cite{hubbard63} 
on the square lattice in two-dimensions 
within the mean-field (BdG-type) approximation:
\begin{eqnarray}
\label{e6}
\Hbdg &=& \hat H_0 + \hat H_\text{I} \\
\hat H_0 &=&
-t\sum_{\langle i,j\rangle , \sigma} \hat{c}_{i,\sigma}^\dagger \hat{c}_{j,\sigma} + \text{h.c.} + 
\sum_{i=1, \sigma}^{\Nbf} \left( V_i-\mu \right) \hat{n}_{i,\sigma} 
\nonumber
\\ 
\hat H_\text{I} &=&-\frac{U}{2} \sum_{i=1, \sigma}^{\Nbf} n(\mathbf{r}_i)  \hat{n}_{i,\sigma} 
- \sum_{i=1}^{\Nbf} \Delta(\mathbf{r}_i) \hat{c}_{i,\uparrow} \hat{c}_{i,\downarrow} + \text{h.c.}, \nonumber
\end{eqnarray}
with local occupation number $n(\mathbf{r}_i)= \sum_\sigma \langle \hat{n}_{i,\sigma} \rangle$ , pairing amplitude 
$\Delta(\mathbf{r}_i)=\langle \hat{c}^\dagger_{i,\downarrow} \hat{c}^\dagger_{i,\uparrow} \rangle$, $U>0$ and random potential $V_i \in \left[-W,W \right]$ drawn from a box distribution.
We employ periodic boundary conditions and work at $T{=}0$; 
the chemical potential $\mu$ is adjusted to fix the 
the filling factor to $n=\sum_{i=1,\sigma}^{\Nbf} \frac{\langle \hat{n}_{i,\sigma} 
\rangle}{\Nbf}=0.875$. 
\footnote{The filling 
factor is chosen in a way to be close to half-filling, which favors a high particle-hole overlap, while avoiding the 
ground state degeneracy of the superconducting state with a charge density wave state at half-filling.}

This Hamiltonian is diagonalized by a Bogoliubov transformation
\eq{
&\gamma_{n,\uparrow}^\dagger=\sum_{i=1}^{\Nbf} \left( u_n(\mathbf{r}_i)\hat{c}^\dagger_{i,\uparrow} + 
v_n(\mathbf{r}_i) \hat{c}_{i, \downarrow} \right)
\\
&\gamma_{n,\downarrow}^\dagger=\sum_{i=1}^{\Nbf} \left( u_n(\mathbf{r}_i)\hat{c}^\dagger_{i,\downarrow} - 
v_n(\mathbf{r}_i) \hat{c}_{i, \uparrow} \right). 
}
The particle and hole wave functions $u_n(\mathbf{r}_i)$ and $v_n(\mathbf{r}_i)$ 
are determined solving the BdG equation
\eq{
 \left(
\begin{array}{cc}
h & \Delta \\
\Delta^* & -h^* 
\end{array}
\right) \left(\begin{array}{cc} u_n(\textbf{r}_i) \\ v_n(\textbf{r}_i) \end{array}\right) = \epsilon_n \left(\begin{array}{cc} u_n(\textbf{r}_i) \\ v_n(\textbf{r}_i) \end{array}\right) ,
\label{e9}
}
where the physical sector corresponds to $\epsilon_n>0$ and
\bea
h u_n(\mathbf{r}_i) &=&-t\sum_\mathbf{\hat{\delta}} u_n(\mathbf{r}_i{+}\mathbf{\hat{\delta}}) \nonumber \\
&&+(V_i{-}\mu-U 
\frac{n(\mathbf{r}_i)}{2}) u_n(\mathbf{r}_i) \label{e10} \\
\Delta u_n(\mathbf{r}_i)&=&\Delta(\mathbf{r}_i) u_n(\mathbf{r}_i);  
\eea
the sum over $\hat \delta$ is over the lattice sites neighboring $\br_i$. 
The scf-conditions for the density $n(\br)$ and the gap-function $\Delta(\br)$ read 
\eq{
&\Delta(\mathbf{r}_i) = U \langle \hat{c}^\dagger_{i,\uparrow} \hat{c}_{i,\downarrow} \rangle = U \sum_n 
u_n(\mathbf{r}_i)v_n^*(\mathbf{r}_i), \label{e12} \\
&n(\mathbf{r}_i)= \sum_\sigma \langle \hat{c}^\dagger_{i,\sigma} \hat{c}_{i,\sigma} \rangle= 2\sum_n|v_n(\mathbf{r}_i)|^2.
\label{e13}}
We assume self-consistency to be attained, if the relative change 
per iteration cycle in $\Delta(\br_i)$ is at each site $\br_i$ 
smaller than $\alpha$. Typical values we take are $\alpha=0.1\%, 0.5\%, 1\%, 3\%$. Note that the average change $\alpha_{\text{avg}}$ per iteration 
cycle is much smaller than $\alpha$, e.g., for a typical sample at moderate disorder 
$W=2$ we have $\alpha_{\text{avg}}=0.014\%, 0.025\%, 0.05\%, 0.1\%$ 

\subsection{Matrix-free implementation of sparse-matrix vector product} 
To speed up a single self-consistency iteration we optimize the Chebyshev expansion. 
Its performance critical part constitutes of the recursive action of the Hamiltonian on a basis vector,
Eq. \eqref{e5}. An implementation of the sparse-matrix vector product custom-tailored to our system is crucial for an optimal performance. 
The sparse-martrix vector multiplication is memory-bound, 
i.e. the performance is limited by the time it takes to fetch data from memory. 
For this reason we devised a self-written "matrix-free" matrix vector product 
that outperforms standard state of the art sparse-matrix vector multiplication libraries.

The idea is the following: Conventional sparse matrix packages keep all non-zero elements, i.e. value and index, in memory. 
Matrix-free implementations become efficient if many of the non-zero elements have identical values storing only the different values that occur. 

With matrix-free implementations the graph of the Hamiltonian has to be hard-coded in the matrix-vector product routine. For our Hamiltonian the amount of memory load operations of matrix data is reduced by a factor of 6 reflecting the number of non-zero elements per row of $\Hbdg$. In addition, the  integer indices corresponding to the matrix graph do not have to be loaded. Altogether, this leads to a reduction of data to be loaded by a factor of 9.
\footnote{The datatype for values is double and for the indices is integer. Note, that the speed-up to be expected from the matrix-free implementation is less 
than a factor of 9. This is because not only the matrix but also the basis vectors have to be loaded from memory, so the reduction of memory load operations also depends on how many basis vectors are acted on in parallel.} 
We mention 
that recently a library has been made available that 
automatizes the implementation of such a matrix-free matrix-vector product for 
a given Hamiltonian \cite{pieper17}.

\subsection{Improved convergence of scf-cycle}
We improve the code performance by reducing the number of iterations needed until
the convergence of the scf-cycle.
The main idea applies, e.g., when scanning the parameter space at fixed $U$ 
for increasing disorder strength $W$. At strength $W_1$ a converged solution $\Psi_1$
is found for a given disorder realization. Thereafter, a sample at larger strength $W_2{>}W_1$ is 
generated by rescaling the disorder in the first sample by a factor of $W_2/W_1$. 
Then, $\Psi_1$ will be used to initialize the scf-cycle for the second sample. 

\subsection{Scaling and Design Considerations}
\begin{figure}[t]
\includegraphics[width=0.5\linewidth]{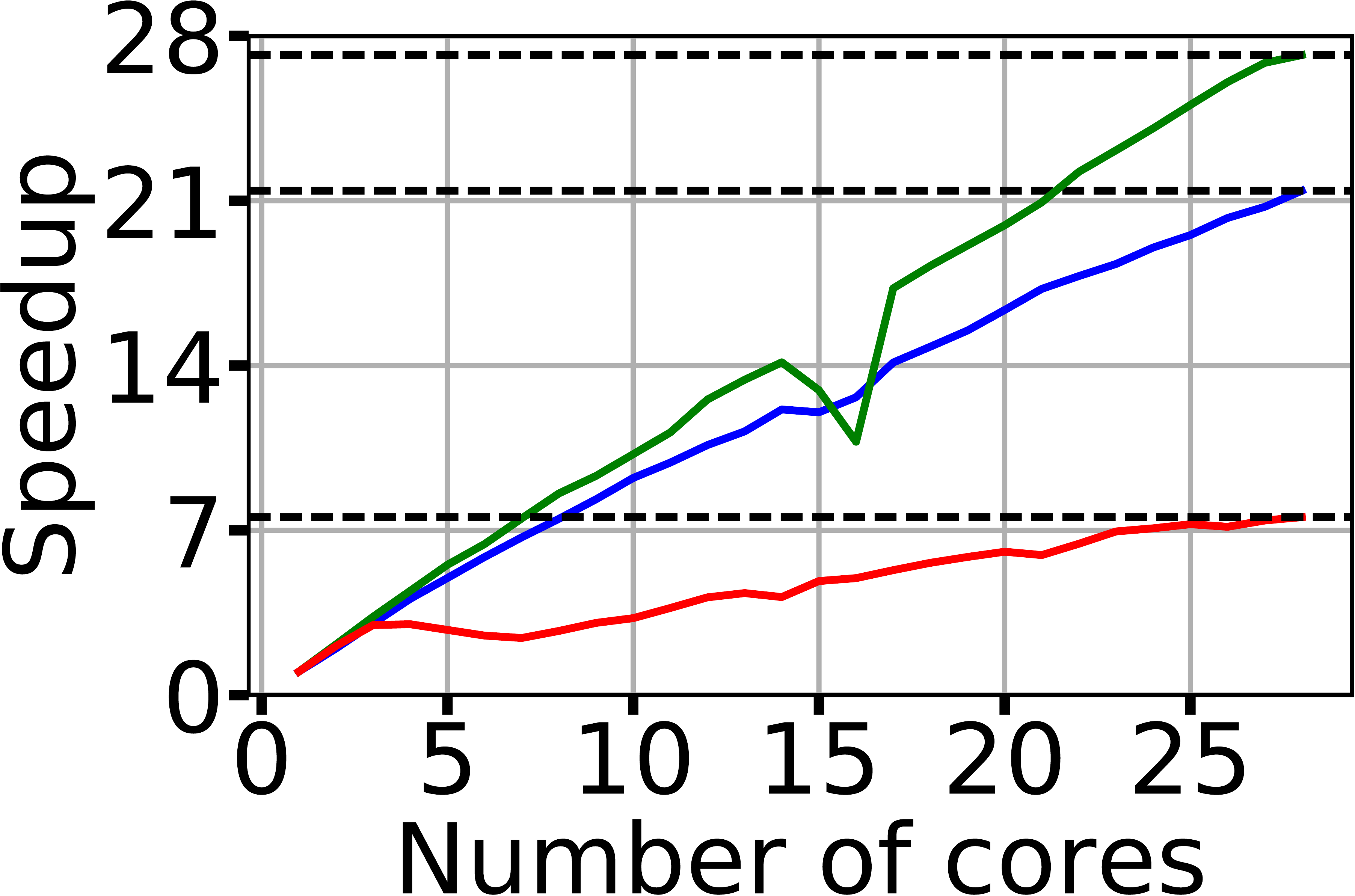}
\includegraphics[width=0.48\linewidth]{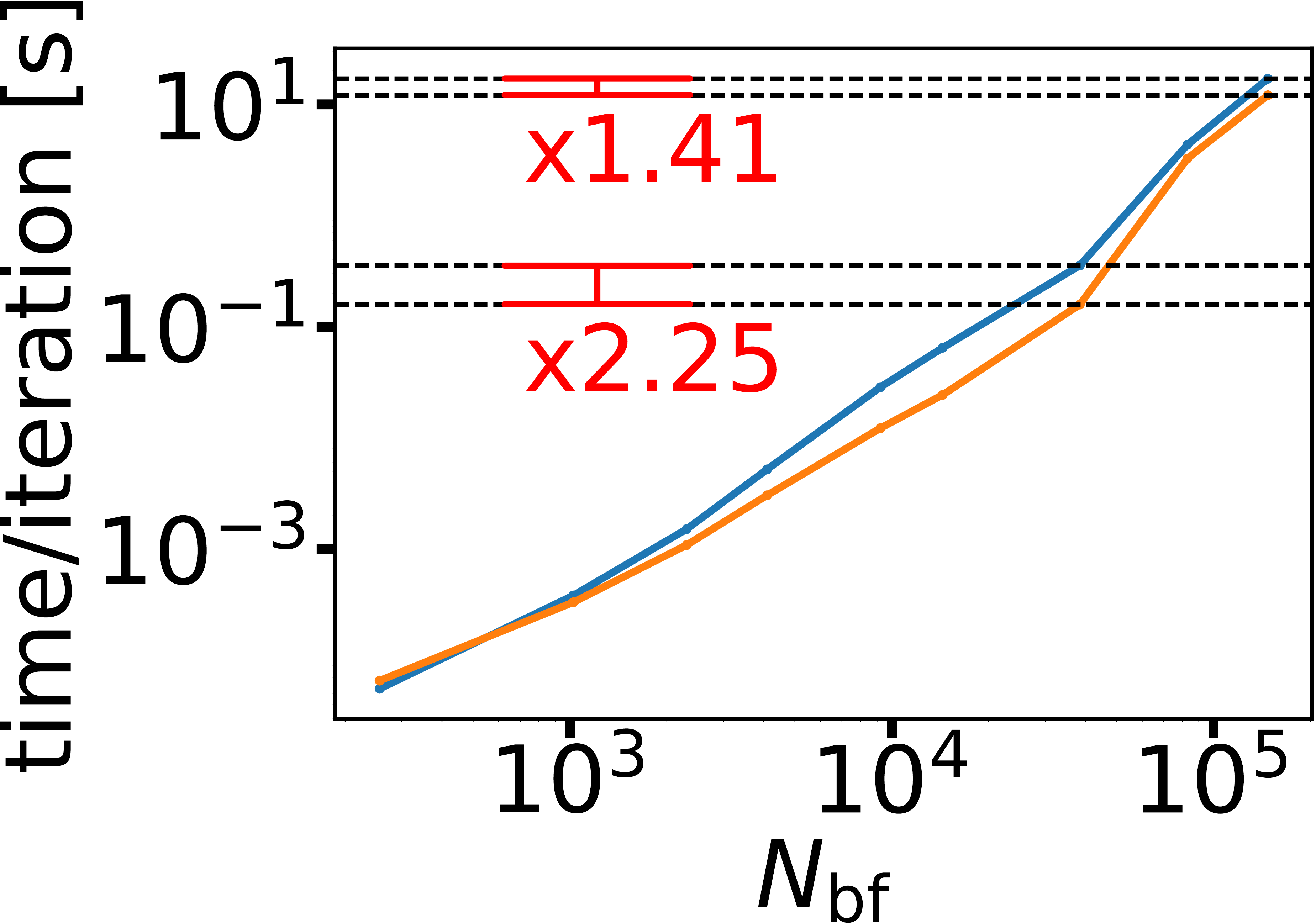}
\caption{Benchmarking intra-node parallelization and code
performance. 
 Left: Speedup with the number of cores per process for different system sizes. The performance dips (green,blue: near 16; red: near 7,14,21,25) with rising number of cores we assign to a hardware issue related to caching. (Parameters: $L=96$ (blue), $L=192$ (green), $L=288$ (red).)
Right: Performance check comparing the matrix-free implementation (orange) with standard 
mkl{\textunderscore}sparse{\textunderscore}d{\textunderscore}mm of the MKL Sparse BLAS library (blue). One iteration corresponds to one sparse matrix-vector product. The ratio of the timings of the MKL and matrix-free algorithms is shown in red at $L=192$ and $L=384$.
\label{f1}
}
\end{figure}

As almost all runtime is spent on the recursive matrix vector products,
 the code lends itself very well to being split in an efficient 
low-level  (i.e. C) kernel embedded in a high-level (i.e. python) 
code that implements the rest of the self-consistency cycle in a 
convenient way with negligible loss of performance. 
The kernel has been optimized for both
threading and vectorization.
In Fig. \ref{f1} we show benchmarks
performed on a compute node 
with two 14-core Haswell Xeon Processor E5-2697 v3; 
we monitor the time spent for performing a single sparse matrix-vector product. 
Fig. \ref{f1} (left) is illustrating  the efficiency of our intra-node (OpenMP) parallelization. For the investigated system sizes $L<288$ the memory-bound runtime limit is not yet reached as is evidenced by the high speedup through parallelization. This makes it very advantageous to perform calculations in this size regime, where parallelization can still be utilized effectively.
Fig. \ref{f1} (right) compares our matrix-free implementation with 
the standard MKL. As is seen from the data, our matrix-free implementation 
is advantageous already at system sizes as small as $N_\text{bf}\sim1000$ sites. 
Note, that at such small system sizes even full diagonalization routines can compete. 
As a technical remark we mention that, in principle, the matrix-free code should always 
be faster as compared to MKL implementations. The crossover size 
originates from our decision to use python as a platform, which leaves an interface to a 
C-based kernel. This interface is plagued with a small overhead that becomes negligible 
beyond the cross-over size.

An additional level of parallelism is obtained by running the expansion of different basis vectors independently on different nodes. The average over the disorder ensemble is performed via farming. This inter-node parallelization scales almost perfectly.

\section{Results: BdG-study of disordered superconductors}

\subsection{Mesoscopic fluctuations  of LDoS and local gap function}
\begin{figure}[t]
\hspace*{-0.3cm}
\includegraphics[width=0.5\linewidth]{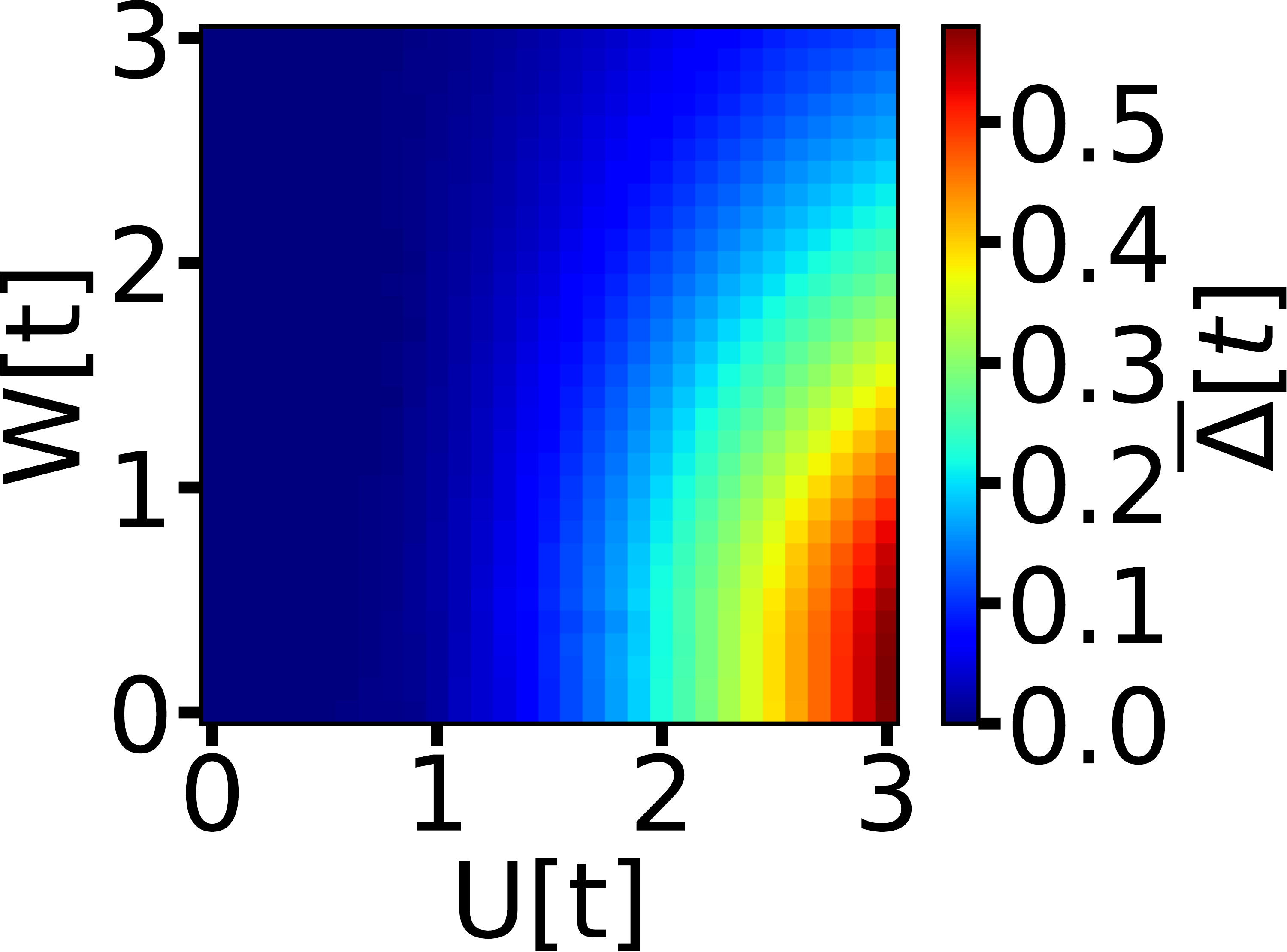}
\includegraphics[width=0.5\linewidth]{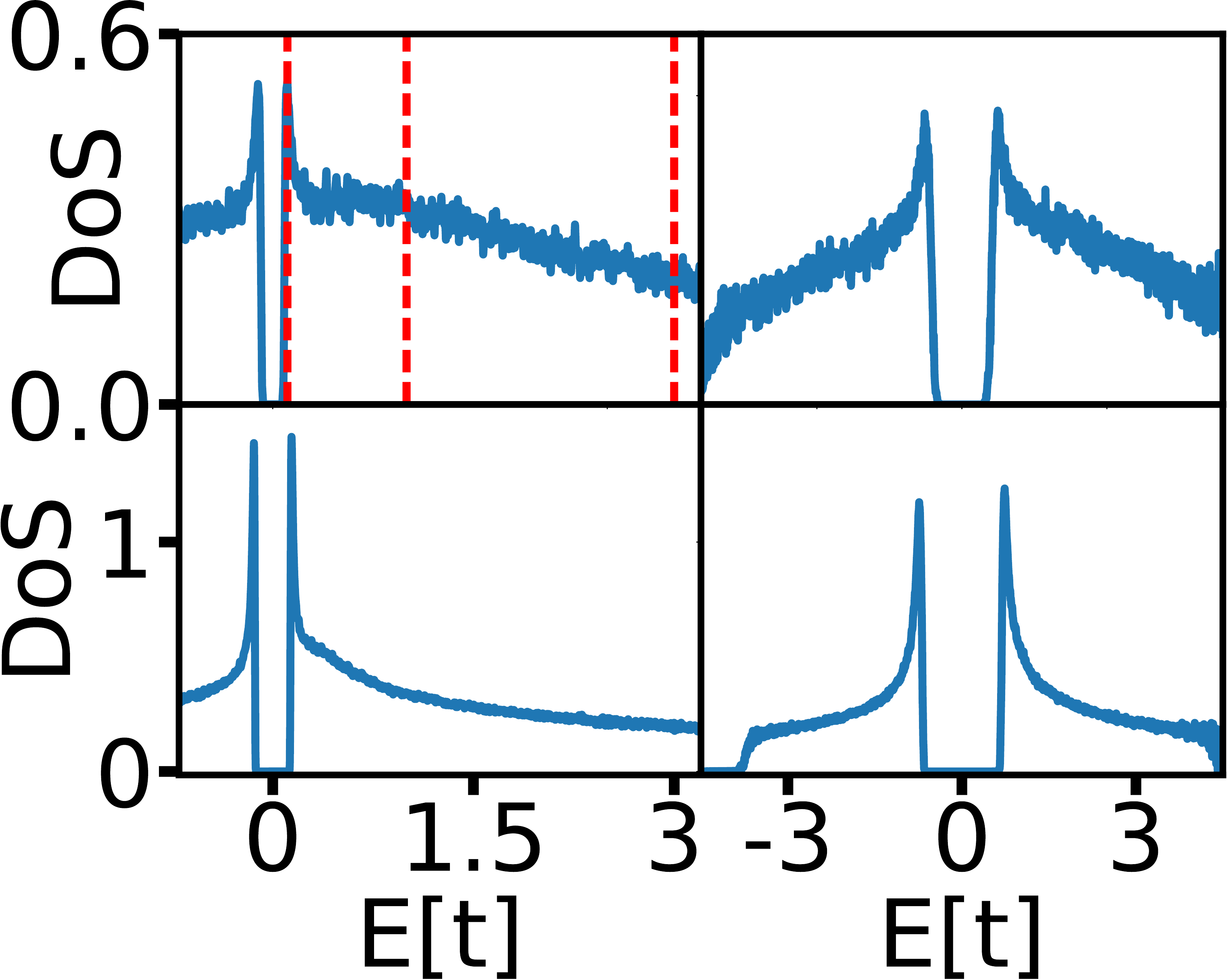}
\caption{Left: Disorder averaged gap $\overline{\Delta(\br)}$ in the $U{-}W$ parameter plane. 
Parameters: $L{=}64$; $\NE=500$, $\alpha=1\%$.
Right: Density of states for typical samples shown at four characteristic points. The red lines indicate the energies at which the LDoS is investigated in Fig. \ref{f4}.
Parameters: $W{=}0.5$ (bottom)  $,1.5$ (top) and $U{=}1.5$ (left) $,3.0$ (right), $L{=}192$; $\Nc=6144$, $\alpha=3\%$.
\label{f3}
}
\end{figure}

As a first application of our technology, we investigate statistical properties of 
$\Delta(\br)$ and of the local 
density of states (LDoS), $\rho(\br,E)$, 
throughout the $U{-}W$-plane. 
To give a first impression we display in Fig. \ref{f3} (left)  
the gap function averaged over a suitable ensemble 
of disordered samples, $\overline{\Delta}(U,W)$; 
the overline indicates the ensemble average with $N_E$ disorder
configurations, typically $\NE \approx 700-800$ samples.
The data has been obtained on a square lattice 
and should be compared with an analogous plot 
produced on the honeycomb lattice by \textcite{Potirniche2014}. 
The gap enhancement seen for the honeycomb lattice is 
not reproduced in Fig. \ref{f3}. This is somewhat surprising, 
perhaps, because the phenomenon on 
the honeycomb lattice has been interpreted in terms of analytical 
results from quantum-field theory\cite{Burmistrov2015ISB}, which also should
apply to the square lattice.

Also displayed in Fig. \ref{f3} (right) is the density of states,
$\rho(E){=}\int_{L^2} d\br \rho(E,\br)$, calculated for four samples in 
representative regions of the parameter plane. 
At weak disorder the spectral gap and the coherence peaks are 
readily identified. Notice that only in the limit of weak disorder
the spectral gap and $\overline\Delta$ scale with each other. \cite{ghosal01}

To characterize the statistical properties of 
physical observables we focus in the following 
on autocorrelation and distribution functions. 
We will compare numerical findings with predictions from analytical theories and, 
in particular, study the sensitivity of qualitative 
results on modifications in the scf-conditions. 
%

\subsubsection{Distribution functions of LDoS and local gaps} 

\begin{figure}[t]
\includegraphics[width=0.49\linewidth]{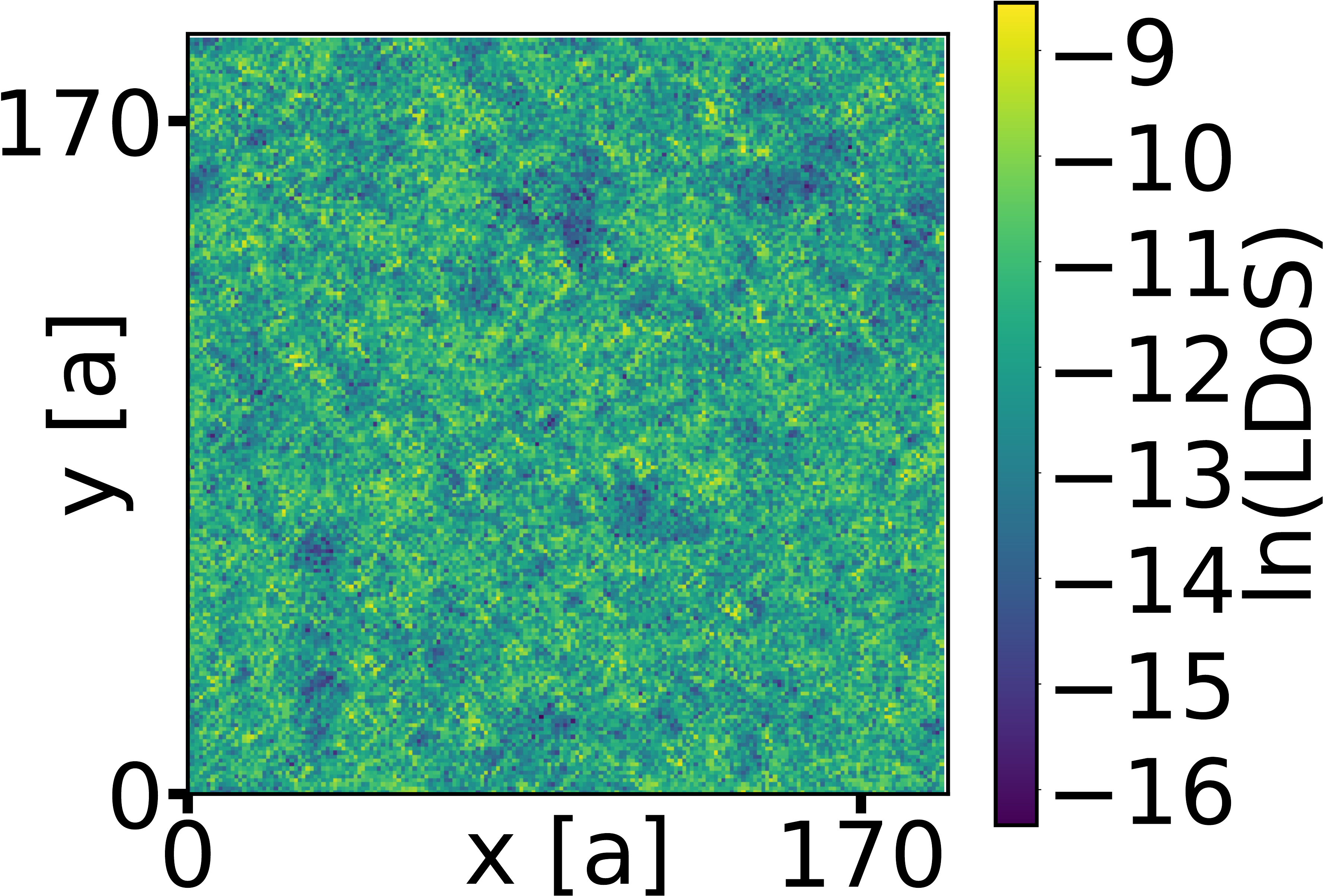}
\includegraphics[width=0.49\linewidth]{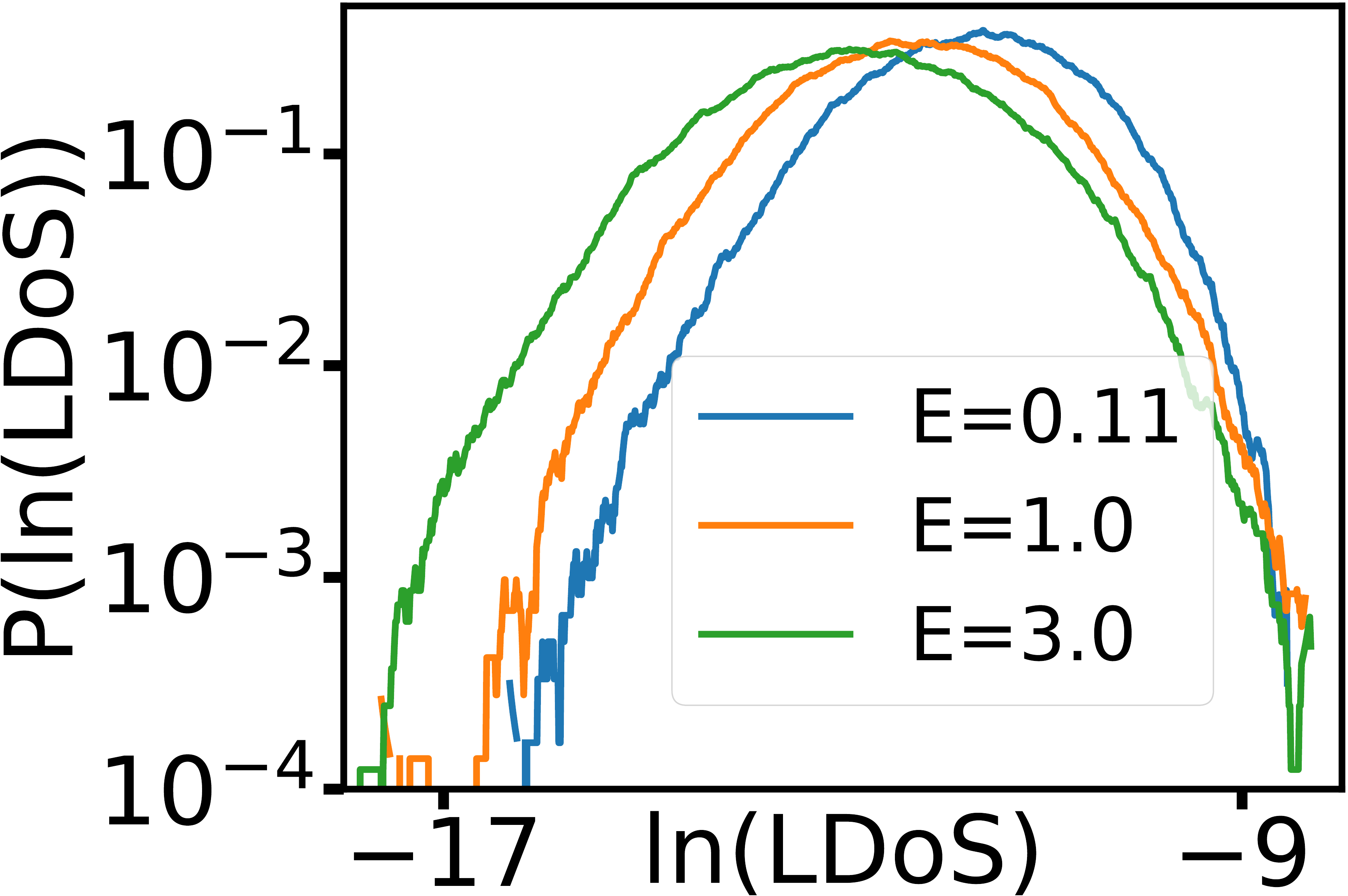}
\caption{Distribution of the local density of states (LDoS), $\rho(E,\br)$. 
Left: spatial distribution for a typical sample at peak energy of DoS
($E{=}0.11$, c.f. Fig. \ref{f3}) 
Right: Corresponding distribution function of LDoS at energies 
$E{=}0.11,1.0,3.0$ illustrating the 
flow of the distribution with $E$. In Fig. \ref{f3} the corresponding DoS can be found.
(Parameters: $W{=}1.5$, $U{=}1.5$; energy resolution $0.013$; $\Nc=6144$, $\alpha=3\%$). 
\label{f4}
}
\end{figure}

\subparagraph*{LDoS.}
We begin the statistical analysis with the spatial fluctuations of 
the LDoS, $\rho(E,\br)$. Fig.\ref{f4} (left) displays an example 
showing how the LDoS is spatially distributed over a typical 
sample with moderate disorder and interaction, $W\gtrsim U\gtrsim 1$.
The logarithmically broad distribution, $\calPld_\rho$,
is readily identified.
The corresponding distribution function 
is displayed in Fig. \ref{f4} (right). 
It takes a log-normal form, already familiar for  
disordered films with size smaller than a localization length,
 see e.g. Eq. (4.101) in Ref. \textcite{Mirlin2000}.  

With increasing energy the distribution shifts to smaller values, 
which is merely reflecting the decrease of the DoS $\rho(E)$, also seen in 
Fig. \ref{f3} (right). 
In contrast, the width of $\rho(E,\br)$ is seen to grow. 
We assign this growths to the fact that the LDoS constitutes an average 
taken over a fixed-size energy window $\eta$. 
The number of eigenfunctions in the averaging window is estimated 
as $\rho(E)\eta L^2$ and therefore changes in energy if the DoS does.
It is larger for energies near the coherence peak as compared to the bulk 
and for that reason the width of $\calPlg$ should be expected to be reduced. 

The LDoS-distribution has been investigated
analytically at temperatures above the critical temperature $T_\text{c}$.  \cite{Burmistrov2016}
Our observations are broadly consistent with these results, since it is reported 
that the distribution develops a pronounced non-Gaussian character upon decreasing 
the temperature. For a more quantitative comparison, simulations at finite temperatures 
are required that are under way. \cite{stosiekUnpublished}

\begin{figure}[b]
\includegraphics[width=0.49\linewidth]{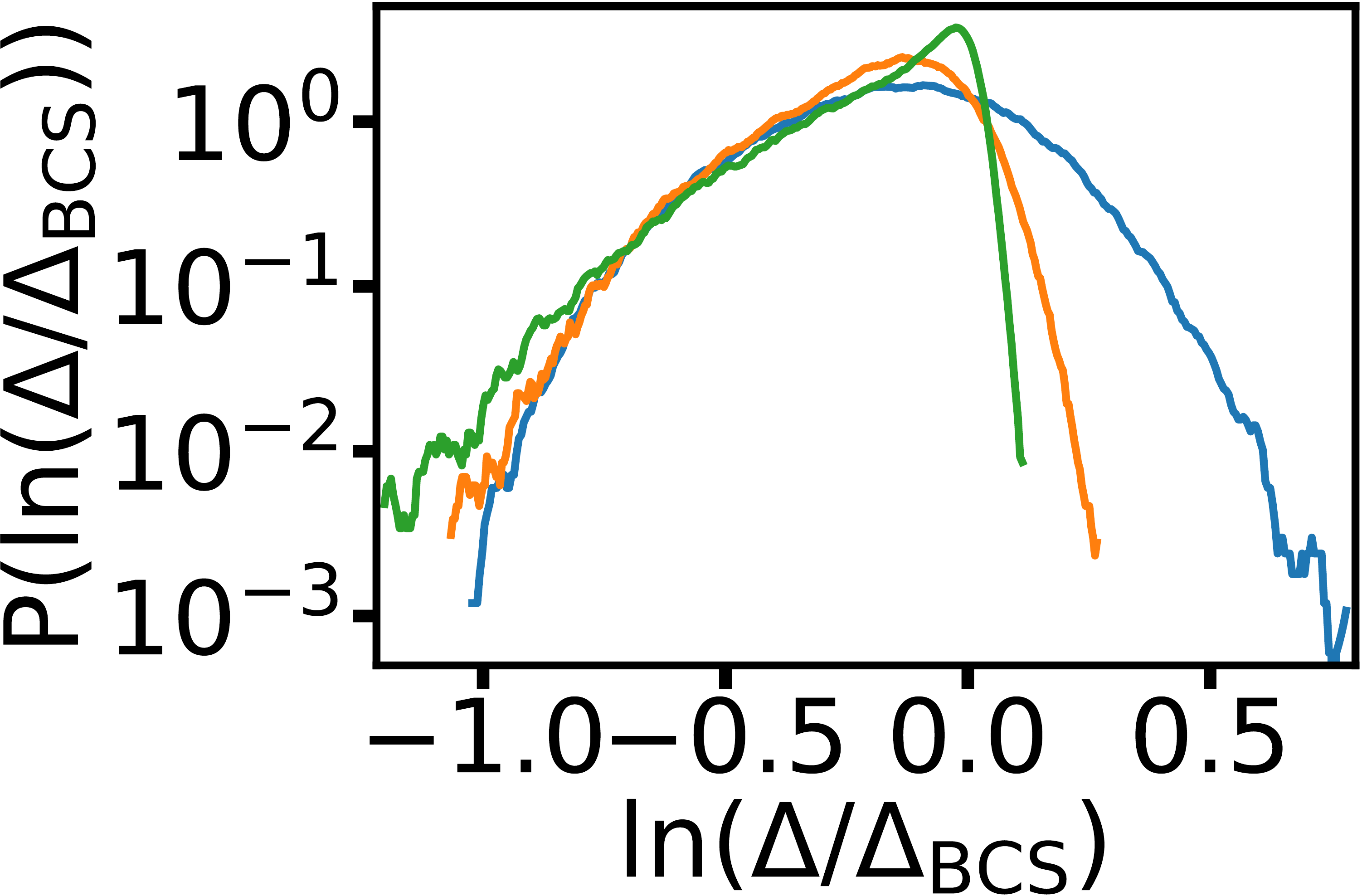}
\includegraphics[width=0.49\linewidth]{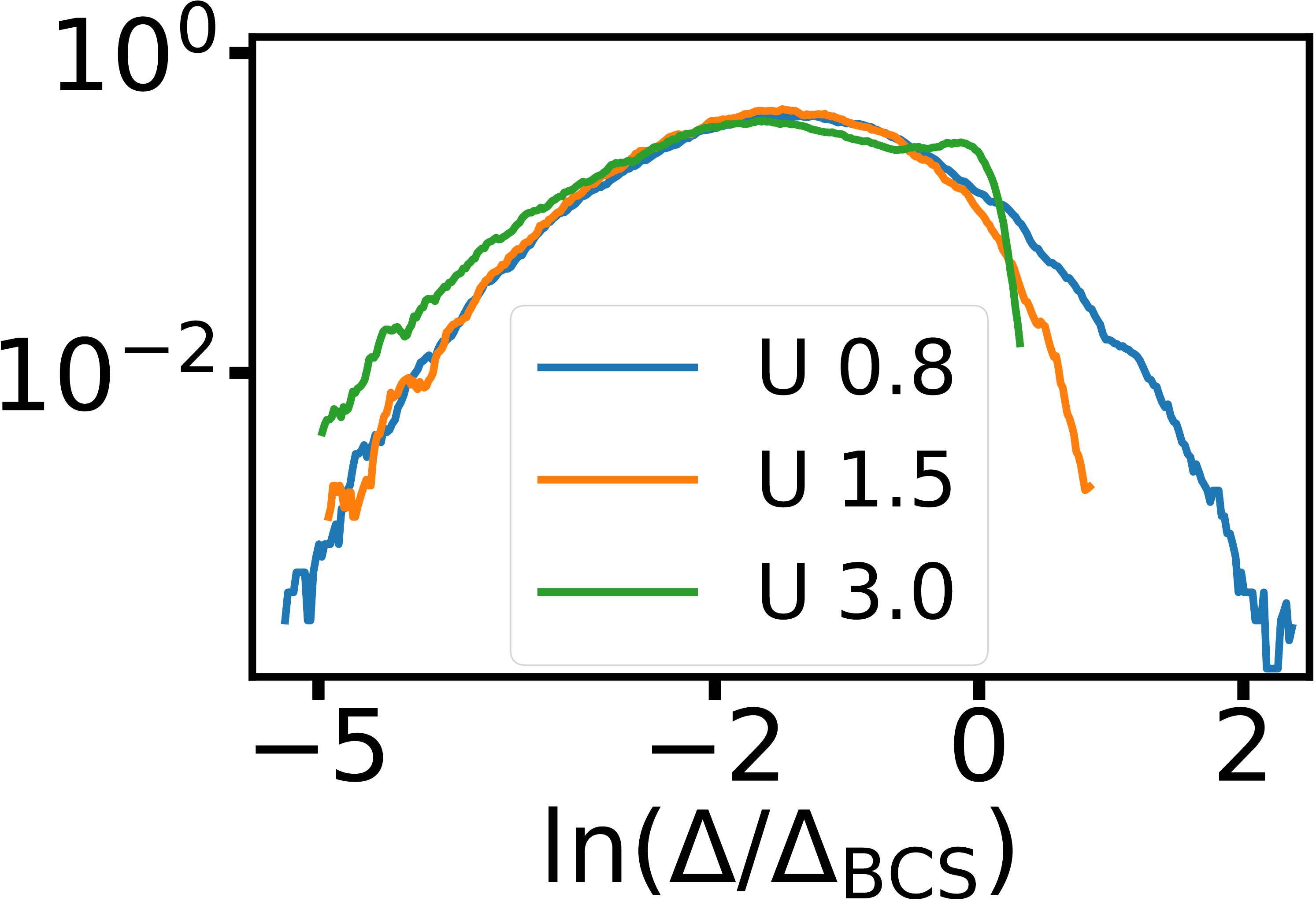}
\caption{Distribution of the local gap function with interaction strength $U$
for a typical sample with $L{=}192$ at weak disorder, $W{=}0.5$ (left plot), 
and stronger disorder, $W{=}2.0$ (right plot). 
As reference energy the pairing amplitude of the clean system, 
$\Delta_\text{BCS}(U)$ has been chosen.
(Parameters: $U{=}0.8,1.5,3.0$; $\Nc=8192,3072,1024$, $\alpha=0.1\%$)
\label{f5}
}
\end{figure}

\subparagraph*{Local order parameter.}
The logarithmically broad distribution of the LDoS is concomitant
with a similarly broad distribution of the local gap function
$\calPld(\Delta)$, Fig. \ref{f5}. 
The evolution of the latter function with interaction strength 
is very interesting. 
As long as disorder, $W$, and interaction, $U$, are weak 
the distribution of the local order parameter is close to Gaussian 
and in this sense roughly following the statistics of the LDoS, 
see Fig. \ref{f5} (left). 
The typical value is seen to be very close to the 
pairing amplitude of the clean system, $\Delta_\text{BCS}(U)$. 
However, with growing $U$ the weight of untypically large values of $\Delta$ 
is seen to be suppressed rapidly, while the weight of 
untypically small values is rather resilient. 

%
%
For increasing disorder and weak $U$ more and more sites 
develop a pairing well below the clean limit, $\Delta(\br)\ll \Delta_\text{BCS}$, 
consistent with observations made in Ref. \cite{ghosal01}. 
Eventually, the shoulder is seen to dominate, 
Fig. \ref{f5} (right) and the distribution $\calPld(\Delta)$ becomes bimodal. 
It features a peak near $\Delta_\text{BCS}$ and a logarithmically distributed background. 
The bimodal shape of $\calPld(\Delta)$ is apparent also from  Ref. \textcite{castellani13} 
where it is seen at very large interaction, $U{=}5$. 

\subsubsection{Autocorrelations of gap function and coherence length}
\begin{figure}[b]
\includegraphics[width=1.0\linewidth]{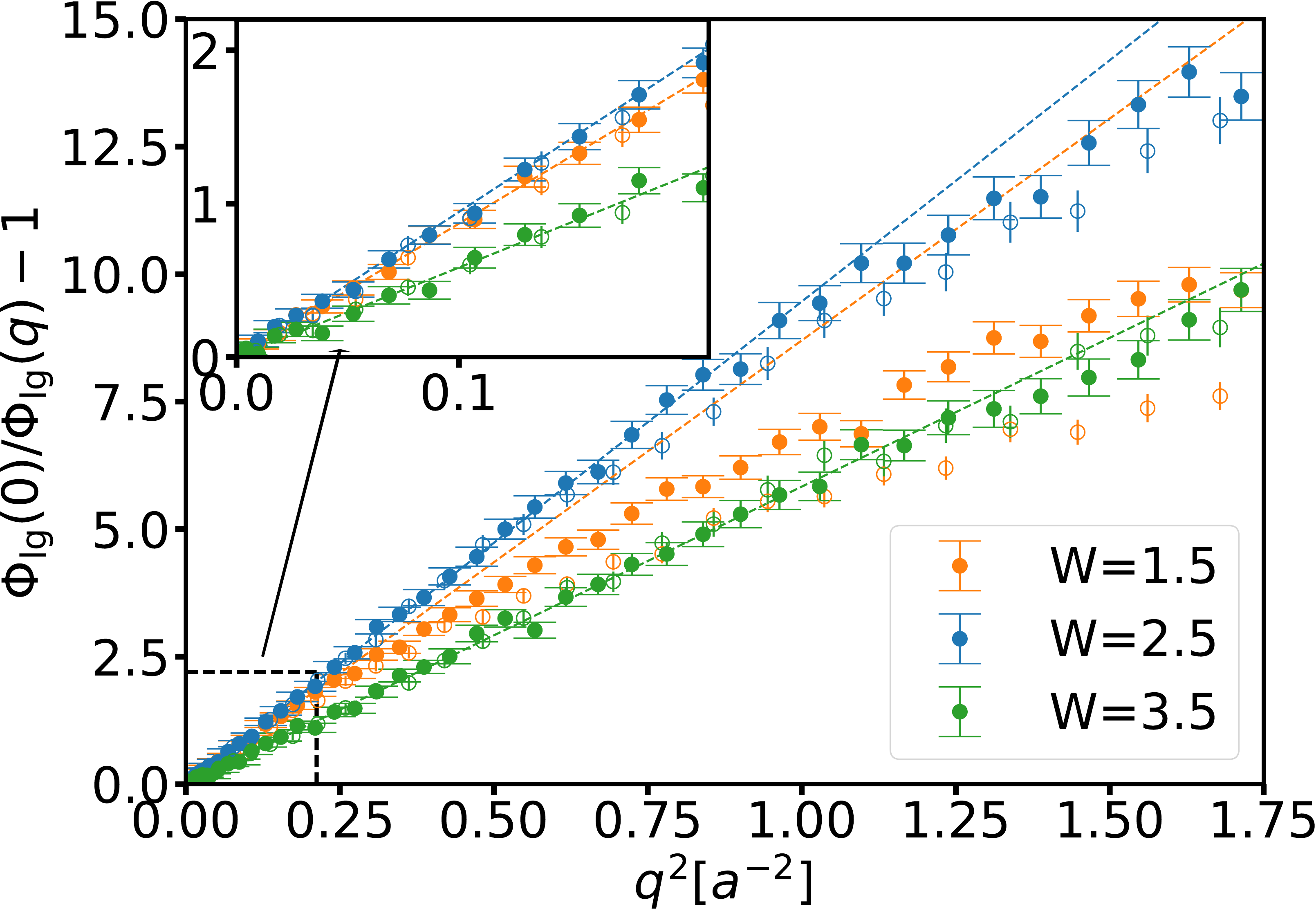}
\caption{The non-trivial part of the inverted normalized gap autocorrelation function $\Philg^{-1}({\bf q}){=}\overline{|\Delta({\bf q})|^{2}}^{-1}$ evolving with $W$ at fixed $U$. $\Philg$ is shown along directions $(\pi/a,0)$ (full symbols) 
and $(\pi/a,\pi/a)$ (open). {\color{black}
The inset shows a blowup of the small wave number regime
where open and closed symbols collapse, so all traces are isotropic. }
(Parameters: $U{=}1.5$, $W{=}1.5$(orange), $2.5$(blue), $3.5$(green) $L{=}192$; $\Nc=1024$, $\NE\approx600-1000$, $\alpha=0.5\%$)
\label{f7}
}
\end{figure}

We consider the disorder averaged spatial autocorrelator   
  $\Philg({\bf q}) = \overline{|\Delta({\bf q})|^2}$ of the pairing function 
  $\Delta(\br)$ in Fourier space.
%
\begin{figure}[t]
\raisebox{-0.025cm}{\includegraphics[width=0.46\linewidth]{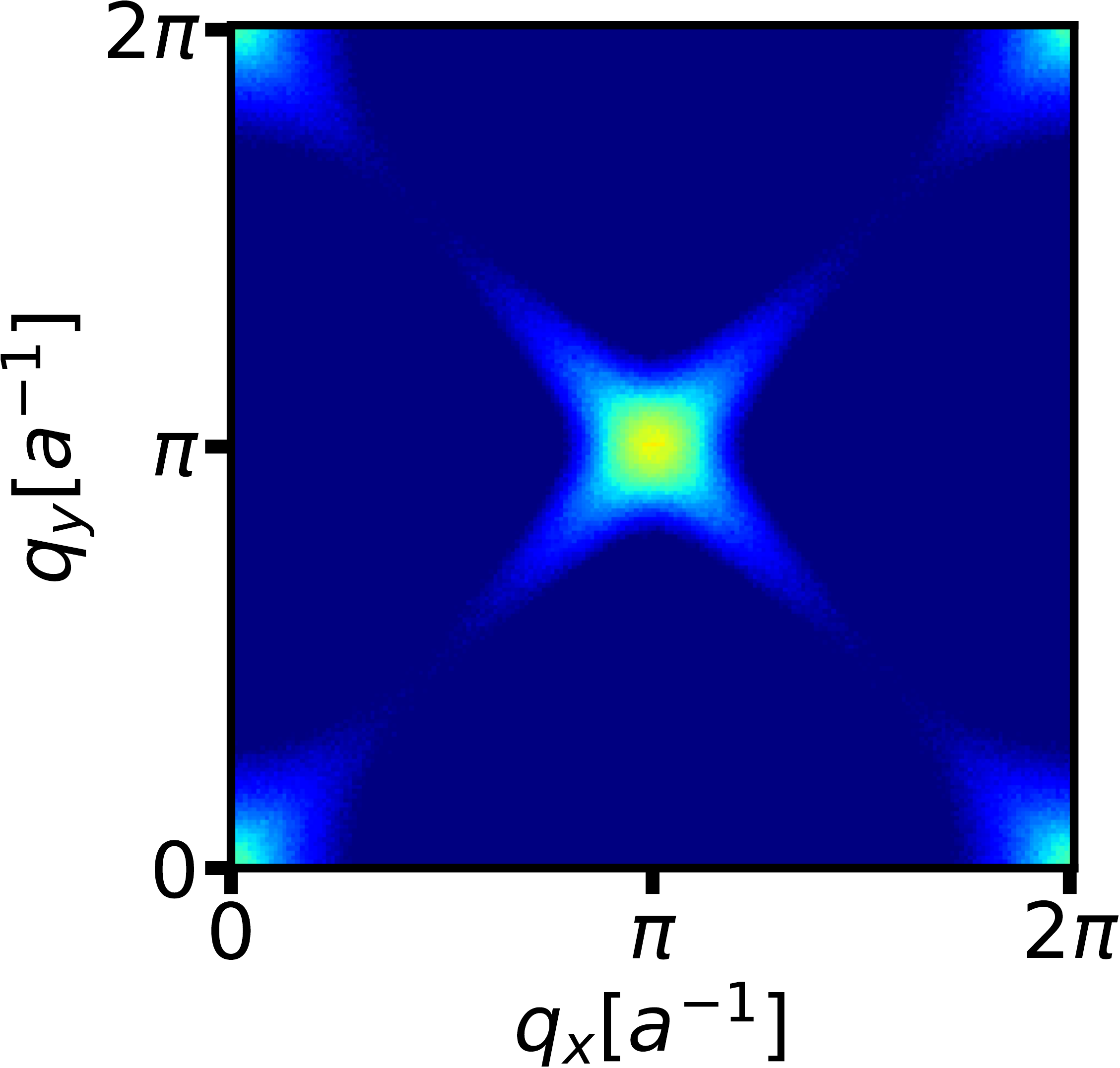}}
\includegraphics[width=0.523\linewidth]{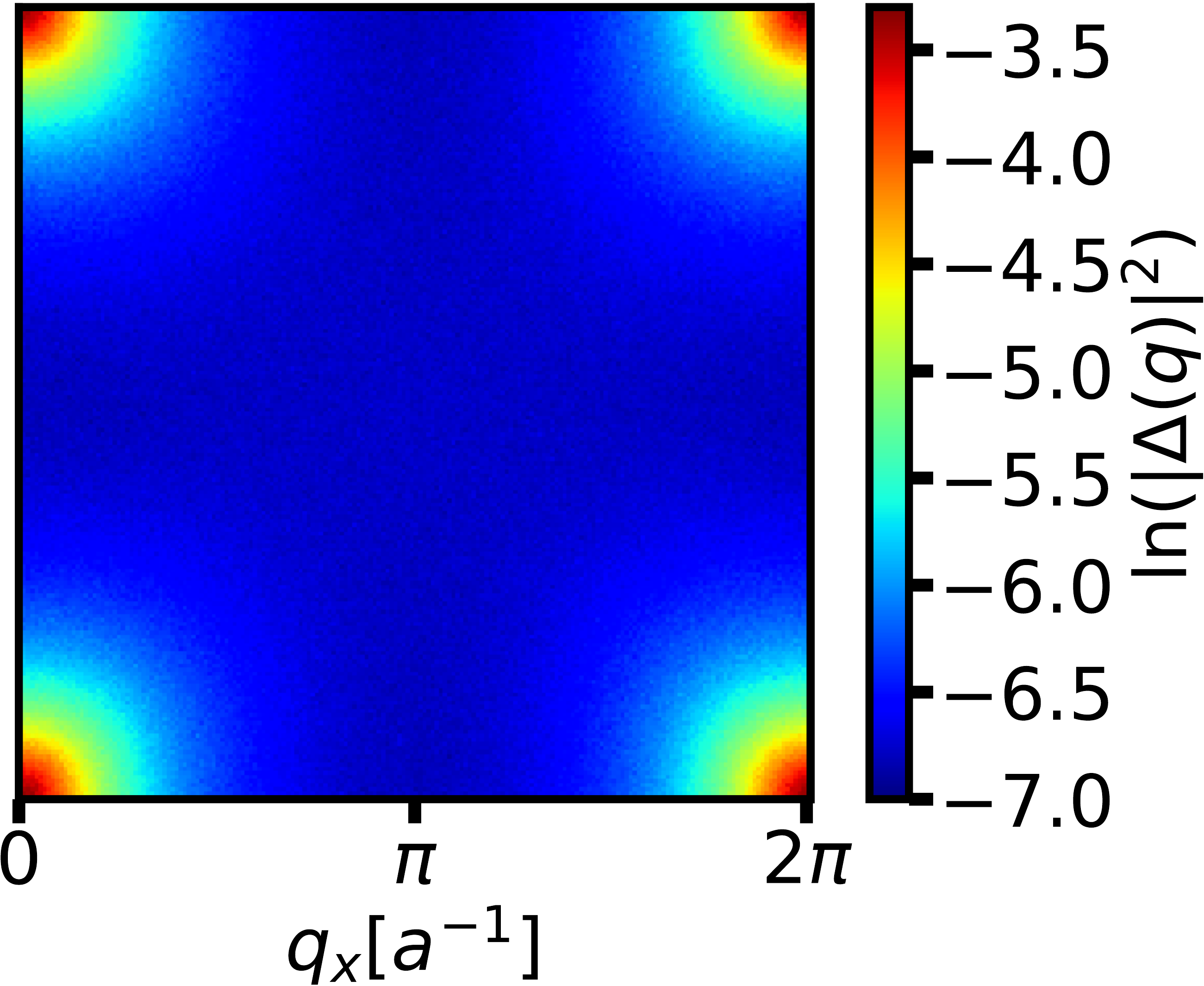}
\caption{The gap autocorrelation function 
$\Philg(q){=}\overline{|\Delta(q)|^2}$ in logarithmic representation
for $L{=}192$ 
and $U{=}1.5$ at two values of disorder, $W{=}0.5$ (left) and 
$W{=}2.5$ (right); $\NE\approx900-1000$, $\alpha=0.1\%$, $\Nc=1024$. 
\label{f6}
}
\end{figure}
At weaker disorder the correlation function
displays a peak at $(\pi/a,\pi/a)$, Fig. \ref{f6}.  
It originates from us choosing the filling fraction $0.875$ 
which is close to the commensurate value unity and thus
should be seen as a signature of the square lattice; 
it disappears at stronger disorder, e.g., at $W{=}2.5$.
The same signature manifests in Fig. \ref{f7} where we show 
$\Philg({\bf q})$ along two directions in ${\bf q}$-space, 
$(\pi/a,0)$ and $(\pi/a,\pi/a)$: As already obvious from Fig. \ref{f6}, 
at wavenumbers of order of the inverse lattice spacing, $a^{-1}$, and
low $W$ the correlator exhibits pronounced deviations from isotropy 
reflected by the collapse of open and closed symbols.

\begin{figure}[b]
\includegraphics[width=0.533\linewidth]{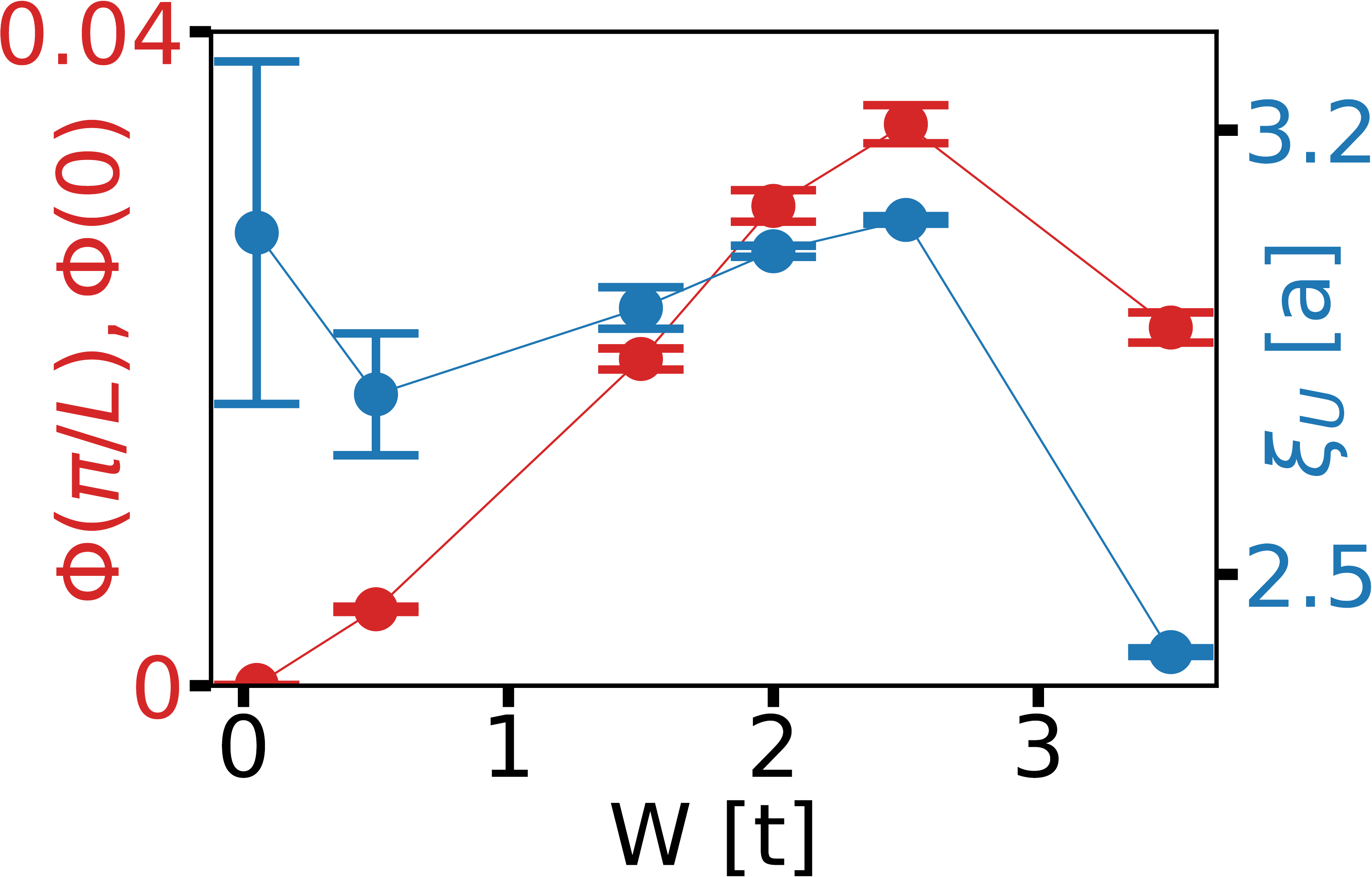}
\includegraphics[width=0.455\linewidth]{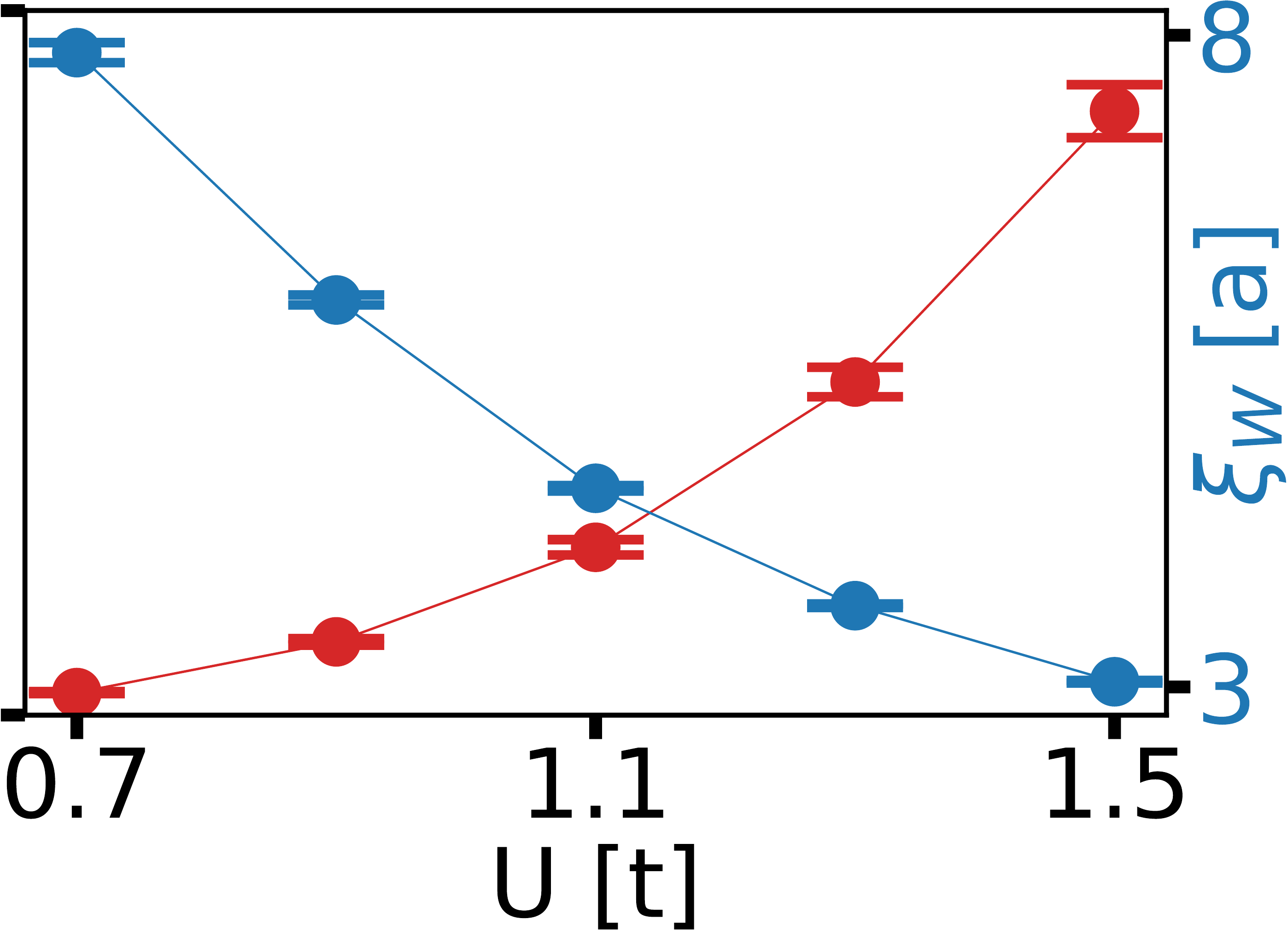} 
\caption{Variation of $\Philg(\pi/L,0)$ and $\Philg(0)$ (red) and the correlation length, $\xi$ (blue) with increasing disorder (left, $U{=}1.5$) and increasing interaction 
(right, $W=2.5$). $\Philg(0)$ coincides with $\Philg(\pi/L)$ within the symbol size as portrayed here. The error bars depict the ensemble average error. The uncertainty due to cutoff $\alpha$ for $\xi_W$ is discussed in the appendix. (Parameters (left): $\Nc=1024$, $\NE\approx600-1000$, $\alpha=0.1\%$. Parameters (right): $\Nc=16384$, $\NE\approx500$, $\alpha=3\%$.)
\label{f8}
}
\end{figure}

Notwithstanding anisotropy at $q{\approx}a^{-1}$, 
in the limit of small wavenumbers 
$q{\ll}a^{-1}$ the correlator $\Philg(q)$ is isotropic and with good accuracy 
we have
\be
\label{e14}
\frac{\Philg(0)}{\Philg(q)} = 1 + (q\xi)^2+\ldots  
\ee
where $\Philg(0)\coloneq\Philg(q\to0)$, given for different $W$ in Fig. \ref{f8}.
$\Philg(\bf q)^{-1}$ behaves nearly quadratically over the whole momentum range where $\Philg(\bf q)$ exhibits isotropic behavior.
Both the increase of $\Philg(0)$ (as approximated by $\Philg(\pi/L,0)$) 
with disorder and the characteristic length $\xi$ have been displayed in Fig. \ref{f8}. 

To attain $\xi$ we have used a linear fit of $\Philg(0)/\Philg(q^2)$ in the isotropic regime. 
As with the range of this regime also the number of data points 
increases considerably with $W$, 
the uncertainty, i.e. the size of the error bars, 
of $\xi$ is seen to decrease with rising $W$ in Fig \ref{f8} (left).
$\xi_U(W)$ exhibits a local non-monotonicity on its way from the clean to the dirty limit; 
the non-monotonic decay is readily seen also from the original data Fig. \ref{f7}.
This peculiar behavior should be interpreted in connection with the formation of superconducting 
islands. It occurs in the same parameter range and may relate to a percolation transition. 
Our data shows that the non-monotonous shape, which was found in 
Ref. \cite{castellani15}
albeit at unrealistically strong interactions $U{=}5$, carries over all the way into the 
physically more relevant regime of intermediate parameter values. 

\section{Impact of self-consistency} 

We return to a central theme of our interest, 
which is the impact of self-consistency on the calculation of physical observables. 

\subsection{Partial (energy-only) self-consistency scheme}

The full BdG-problem is specified by the set of equations 
\eqref{e9} - \eqref{e13}. It is highly complicated, e.g., 
because the scf-conditions \eqref{e12} and \eqref{e13} are 
non-linear. As is frequently done in such situations, 
the full scf-problem is replaced by a simplified variant
exhibiting partial self-consistency. 

Various possibilities for such simplifications are conceivable. 
The scheme we here describe is inspired by analytical calculations
performed by Feigelman et al. \cite{Feigelman2007ISB, feigelman10}.  
The overall procedure can be considered a generalization of BCS 
theory that allows for an inhomogeneous order parameter. 
To bring the self-consistency requirement into the familiar BCS 
form, additional approximations besides the mean-field decoupling 
are necessary.

We here derive equations for partial self-consistency
starting from the mean-field Hamiltonian Eq. \eqref{e6}. 
We express the field operators employing as a basis the eigenstates
$\psi_l(\br_i)$ of the non-interacting part of $\Hbdg$, i.e. $\hat H_0$: 
\eq{ 
\hat{d}_{l,\sigma}=\sum_{i=1}^{\Nbf} \hat{c}_{i,\sigma}\psi^*_l(\mathbf{r}_i), \quad 
\hat{d}^\dagger_{l,\sigma}=\sum_{i=1}^{\Nbf} \hat{c}^\dagger_{i,\sigma}\psi_l(\mathbf{r}_i). 
}
The corresponding eigenvalues of $\psi_l$ are denoted $\xi_l$ and will 
be measured with respect to the Fermi-energy $E_\text{F}$. 
Expressing $\Hbdg$ in $\hat d, \hat d^\dagger$ we obtain 
\eq{
\Hbdg=&\sum_{l=1, \sigma}^{\Nbf} \xi_l \hat{d}^\dagger_{l,\sigma} \hat{d}_{l,\sigma} + \sum_{l,m,n,o, 
\sigma}M_{lmno} \langle \hat{d}^\dagger_{l,\sigma} \hat{d}_{n,\sigma} \rangle \hat{d}^\dagger_{m,\sigma} 
\hat{d}_{o,\sigma} \\
 &- U \sum_{l,m,n,o=1}^{\Nbf} M_{lmno} \langle \hat{d}_{n, \downarrow} \hat{d}_{o, \uparrow} \rangle 
\hat{d}^\dagger_{l,\uparrow} \hat{d}^\dagger_{m,\downarrow} + \text{h.c.} \nonumber,
}
where an abbreviation 
\eq{ 
& M_{lmno}=\sum_i \psi_l^*(\mathbf{r}_i) \psi_m^*(\mathbf{r}_i) \psi_n(\mathbf{r}_i) \psi_o(\mathbf{r}_i),
}
has been introduced. 

The main approximate step in partial self-consistency is to neglect all terms with more than two indices
\eq{M_{lmno}=
\begin{cases}
    M_{ln},& \text{if } l=m \text{ and } n=o\\
    0,              & \text{otherwise}
\end{cases},
}
together with the Hartree term. The simplified mean-field Hamiltonian then reads
\eq{
\label{e18}
\Hbdg^\text{s}=\sum_{l=1,\sigma}^{\Nbf}\xi_l \hat{d}_{l,\sigma}^\dagger 
\hat{d}_{l,\sigma}+\sum^{\Nbf}_{l=1} \Delta_l \hat{d}^\dagger_{l,\uparrow} \hat{d}^\dagger_{l,\downarrow}+\text{h.c.},
}
with an s-wave pairing strength 
\eq{
\Delta_l=-U\sum_{m=1}^{\Nbf}M_{lm}\langle \hat{d}_{m,\uparrow} \hat{d}_{m,\downarrow} \rangle.
}
The Hamiltonian \eqref{e18} is structurally equivalent to the BCS Hamiltonian in the sense 
that the kinetic term and $\Delta_l$ are diagonal in the same (real-space) basis;
Cooper pairs form within a Kramer's doublet. 
The corresponding BCS gap-equation reads 
\eq{
	\Delta_l=\frac{U}{2}\sum_{m=1}^{\Nbf} M_{lm} \frac{\Delta_m}{\sqrt{\Delta_m^2+\xi_m^2}}.
}
Converting back to real-space we have 
\eq{
	\Delta(\mathbf{r}_i)=\frac{U}{2}\sum_{l=1}^{\Nbf} \frac{\Delta_l}{\sqrt{\Delta_l^2+\xi_l^2}} 
\psi_l^2(\mathbf{r}_i).
}
The advantage of the partial (``energy-only'') 
scf-scheme is that the pairing-amplitude can be calculated 
solely from the eigenstates and eigenvalues of  
the non-interacting reference Hamiltonian $ \hat H_0$. 
This comes at the expense of ignoring changes 
in the wavefunctions related to pairing and the inhomogeneous 
Hartree shift.

\subsection{Effects of self-consistency schemes on the local-gap distribution}
We compare the results of full and energy-only 
self-consistency schemes for the local pairing amplitude $\Delta(\br)$ for the Anderson Problem in 2D and 3D.
\subsubsection*{2D}
\begin{figure}[h]
\includegraphics[width=1.0\linewidth]{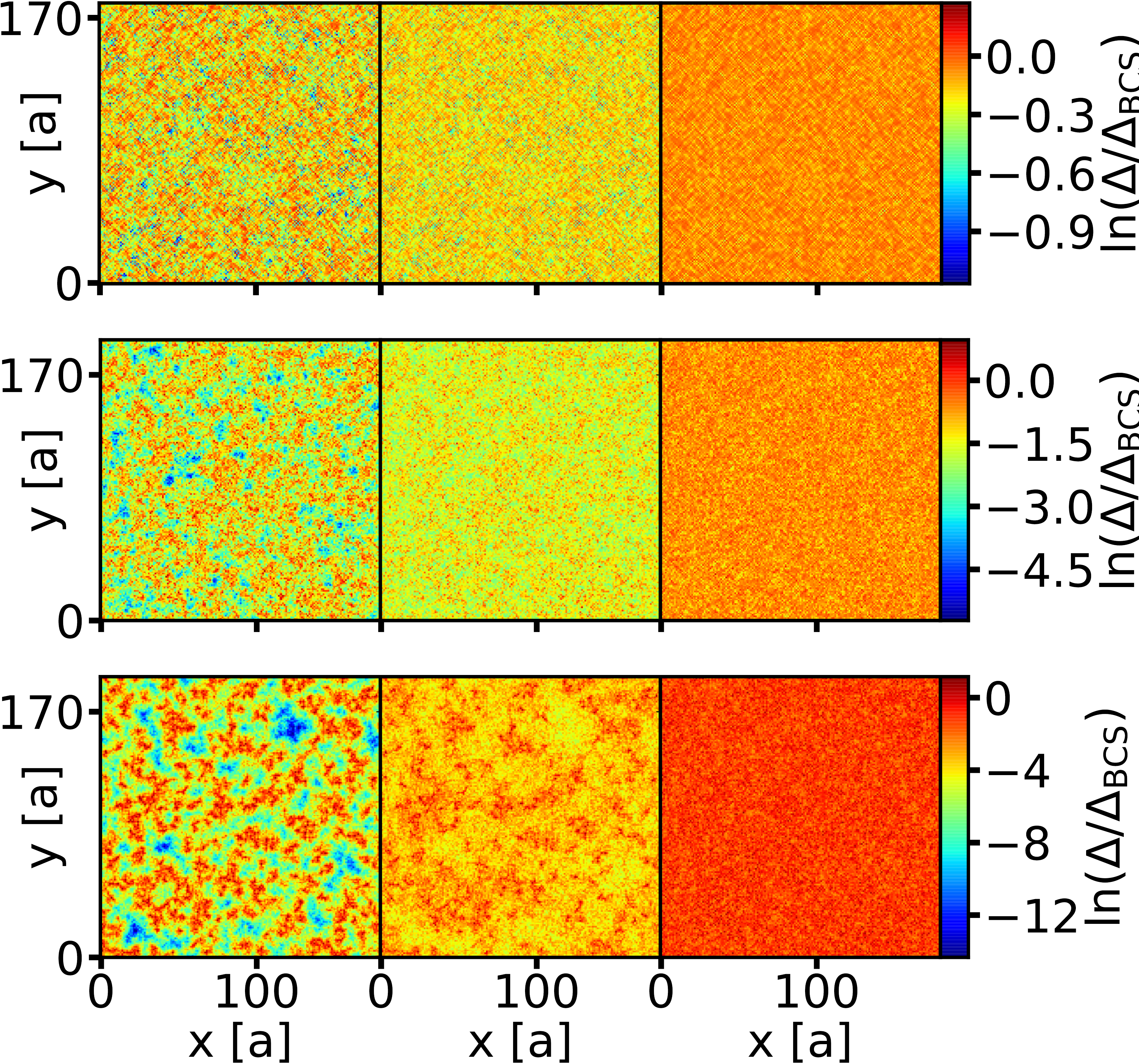}
\caption{Evolution of islands with disorder increasing from top to bottom, 
$W{=}0.5,2.0,3.5$. Different self-consistency schemes are compared. 
Left column: full self-consistency. 
Center column: energy-only self-consistency with inhomogeneous Hartree shift. 
Calculation is done with the single-particle ("screened") potential as it results from the 
full scf-calculation, left. 
Right column: energy-only scheme. The energy-only data has been calculated employing full diagonalization (Parameters: $U{=}1.5$; $\Nc{=}1024$, $\alpha=0.5\%$)}
\label{f9}
\end{figure}
Fig. \ref{f9} shows a spatial distribution 
of $\Delta(\br)$ as obtained for 
typical sample at intermediate interaction and three 
disorder values. 
The calculation with full self-consistency, 
Fig. \ref{f9} (left) column exhibits a clear 
tendency towards the formation of superconducting islands. 
In contrast, with energy-only self-consistency, right column, 
a rather homogeneous speckle pattern is found missing 
any indications of island formation. 
Hence, already by inspecting individual samples 
we expect that distribution functions of physical observables 
will depend in a qualitative way on the applied scf-scheme 
in broad parameter regions. 

In order to highlight the effect of screening, we have displayed in Fig. \ref{f9} also 
the results of an intermediate scf-scheme. It operates in an energy-only mode, 
but adopts for the disorder the effective single particle potential 
("screened" potential) as it is obtained as a 
result from the full scf-calculation. As is seen from Fig. \ref{f9}, center column 
first indications of islands emerge, but the contrast is still largely underestimated. 
This result underlines the importance of full consistency in the scf-procedure.
\subsubsection*{3D}
\begin{figure}[t]
\includegraphics[scale=0.58]{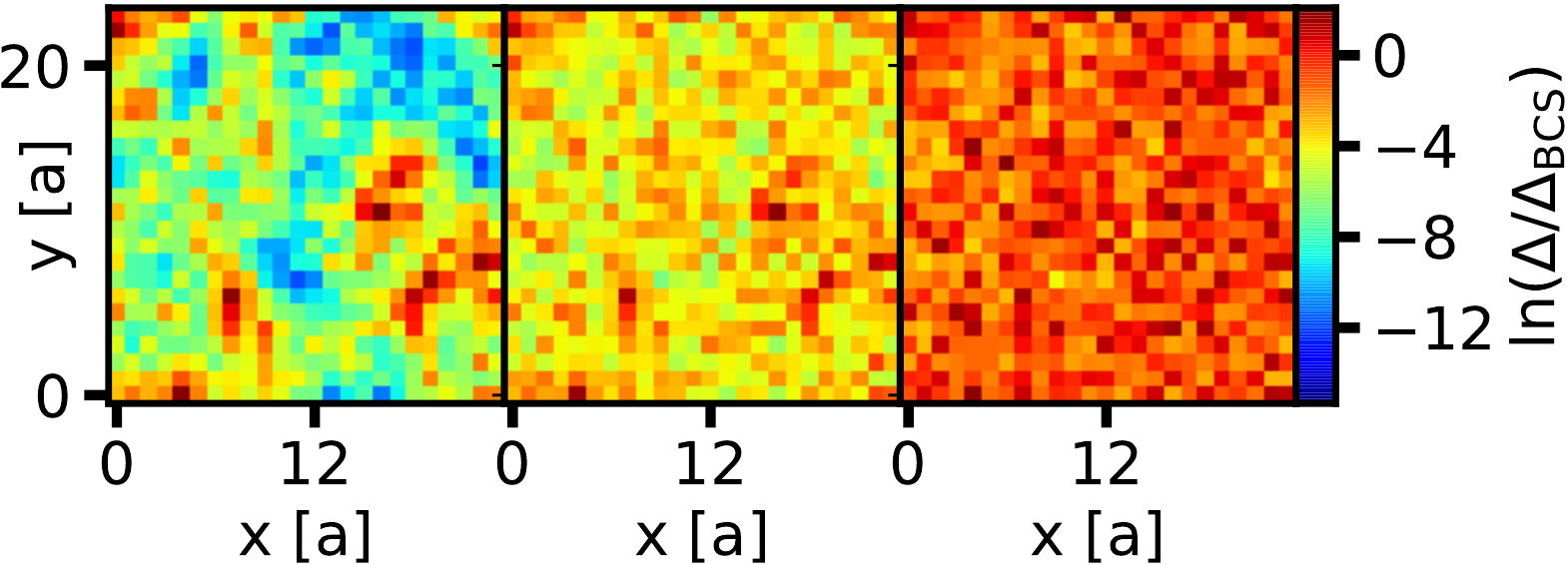}
\caption{Corresponding plot to Fig. \ref{f9} in 3D. A representative 2D slice of a sample is shown. The data has been calculated employing full diagonalization (Parameters: $L=24$ $W{=}4.0$, $U{=}2.5$, $n=0.3$; $\alpha=0.5\%$) \label{f12}}
\end{figure}
In analogy to the 2D case, we compare the results of full and energy-only self-consistency 
schemes for the local pairing amplitude $\Delta(\br)$ in 3D.

In the non-interacting 
3D Anderson problem there is a phase transition at a critical disorder strength $W_c$, where all states become localized. For a disorder strength below $W_c$ there exists an energy $E_c$, 
the mobility edge, which separates a fully localized band from a band of extended states. 
We note that as the Anderson Hamiltonian is symmetric around $E=0$ this is 
also true for the mobility edge. We refer to \textcite{kramer87} for the phase diagram.

Our interest is in the importance of self-consistency in the presence of attractive 
on-site interactions close to the mobility edge in the insulating band.
For comparibility with authors that have considered an energy-only approach in this context before\cite{feigelman10}, we choose a Gaussian disorder distribution
\begin{equation}
	p(V_i)=\frac{1}{\sqrt{2\pi}W}\text{exp}\left[-\frac{V_i^2}{2W^2}\right]
\end{equation}
of the random onsite energies $V_i$ in Eq. \ref{e10}.

Fig. \ref{f12} shows the spatial distribution of $\Delta(\br)$ of a typical sample 
as obtained for moderate interaction and disorder strength and chemical potential in the
localized band. The chosen filling factor $n{=}0.3$ corresponds to a chemical potential of $\mu\approx-6$ 
in the fully self-consistent case. The mobility edge without interactions is located at $E_c\approx-5.5$
for the disorder strength $W{=}4$ that is considered here\cite{kramer87}. As in the 2D case, the field obtained within the fully self-consistent calculation shows a pronounced formation of islands, Fig. \ref{f12} (left). The energy-only scheme in analogy to our 2D results exhibits a rather homogeneous spatial distribution, Fig. \ref{f12} (right). The results of the energy-only scheme with "screened" potential shown in Fig. \ref{f12} (center) again show first indications of island development with dramatically underestimated contrast. 
This highlights the importance of full self-consistency also in 3D. 

To what extent the conclusions of earlier theoretical works that consider 
this scenario \cite{Feigelman2007ISB, feigelman10} are affected 
remains to be seen.

\subsection{Effects of self-consistency on gap autocorrelator}
\begin{figure}[t]
\begin{center}
\includegraphics[width=1.0\linewidth]{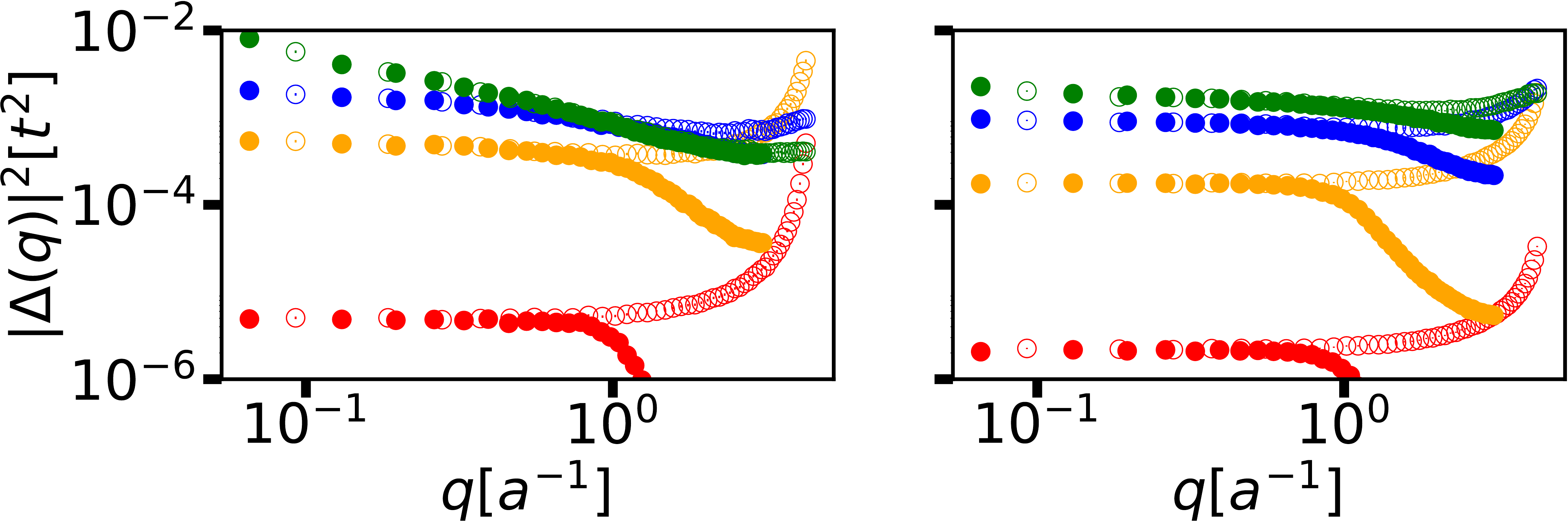}
\end{center}
\caption{
Gap autorcorrelation function $\Philg({\bf q}){=}\overline{|\Delta({\bf q})|^2}$ calculated employing 
two different energy-only self-consistency schemes. 
$\Philg$ is shown along directions $(\pi/a,0)$ (full symbols) 
and $(\pi/a,\pi/a)$ (open); traces for four different disorder values 
are shown, $W{=}0.05,0.5,1.5,2.5$, from bottom to top. 
Left: energy-only self-consistency with screened potential.  
Right: energy-only self-consistency.
(Parameters: $U{=}1.5$ $L{=}96$; $\NE=1000$, $\alpha=0.1\%$)
\label{f10}
}
\end{figure}

Fig. \ref{f10} shows data analogue to Fig. \ref{f7}, 
now with energy-only self-consistencies. 
As is obvious already from individual sample,
Fig. \ref{f9}, the contrast parametrized by $\Philg(0)$
is much smaller as compared to the case of full 
self-consistency given in Fig. \ref{f6}. 
As one would expect from Fig. \ref{f9}, 
the contrast with screened potential, Fig. \ref{f9} (right)
exceeds the bare scheme, Fig. \ref{f9} (left) 
considerably. 

The most striking and perhaps unexpected feature, however, is a qualitative difference. 
In the full scf-calculation, $\Philg(q)$ follows Eq. \eqref{e14} and 
exhibits a well defined parabolic shape in the vicinity of small wavenumbers. 
This feature is not reproduced within energy-only schemes.  
The bare scheme does not exhibit an appreciable curvature 
up to $q\approx a^{-1}$, so 
$\xi_\text{eo} \approx 0$.
In contrast, within the screened scheme $\calPlg$ 
does not show clear saturation at small wavenumbers 
within the range of $q$-values accessible. 
We thus interpret these results as a strong indications that 
wavefunction renormalization as it occurs within the full scf-scheme 
is crucial for understanding those aspects of qualitative physics 
that hinge on long-range spatial correlations.

\section{Conclusions}
We have implemented an efficient solver of self-consistent field (scf-) Hamiltonians
that is based on the kernel-polynomial method. An application to disordered 
$s$-wave superconducting films has been presented that employs the Bogulubov-deGennes
approximation. The statistical properties of the local density of states and of the 
local gap function $\Delta(\br)$ have been studied. 
In this context our computational machinery proves useful since 
system sizes can be accessed significantly 
exceeding the ones that have been achieved in the earlier work. 
We thus can study the crossover in disorder strength $W$ and interaction strength $U$
from the strongly coupled into the perturbative regime, where analytical methods apply 
and can provide conceptual insights. 

Along this way three key observations have been made. 
(i) Superconducting islands form in large regions of the $U{-}W$ phase space 
and thus appear to be a typical encounter already at intermediate interaction and disorder strength. 
(ii) Presumably related to island formation, the (mean-field) correlation length 
exhibits a non-monotonous variation when sweeping from very weak to strong disorder. 
(iii) Island formation is a hallmark of wavefunction-renormalization in the sense that 
islands do not form with partial ("energy-only") self-consistency. 
To investigate into possible consequences of this observation  for analytical treatments of the 
superconductor-insulator transition we leave as a topic for future research. 

As a concluding remark we note that the BdG-Anderson problem and the associated 
ensemble of self-consistent random Hamiltonians is a particular representative of 
a very large class of random matrices that satisfy a self-consistency constraint 
("scf-ensembles"). Presumably, because of the considerable challenges 
that such ensembles imply for analytical and computational treatments very little 
is known about them. We take the observations that have been reported  
for the BdG-ensembles, in this work as well as by the earlier authors, 
as a strong indication that much is there to be discovered. 

\section*{Appendix}
\subsection*{Self-consistency cutoff discussion}
\begin{figure}[t]
\includegraphics[width=1.0\linewidth]{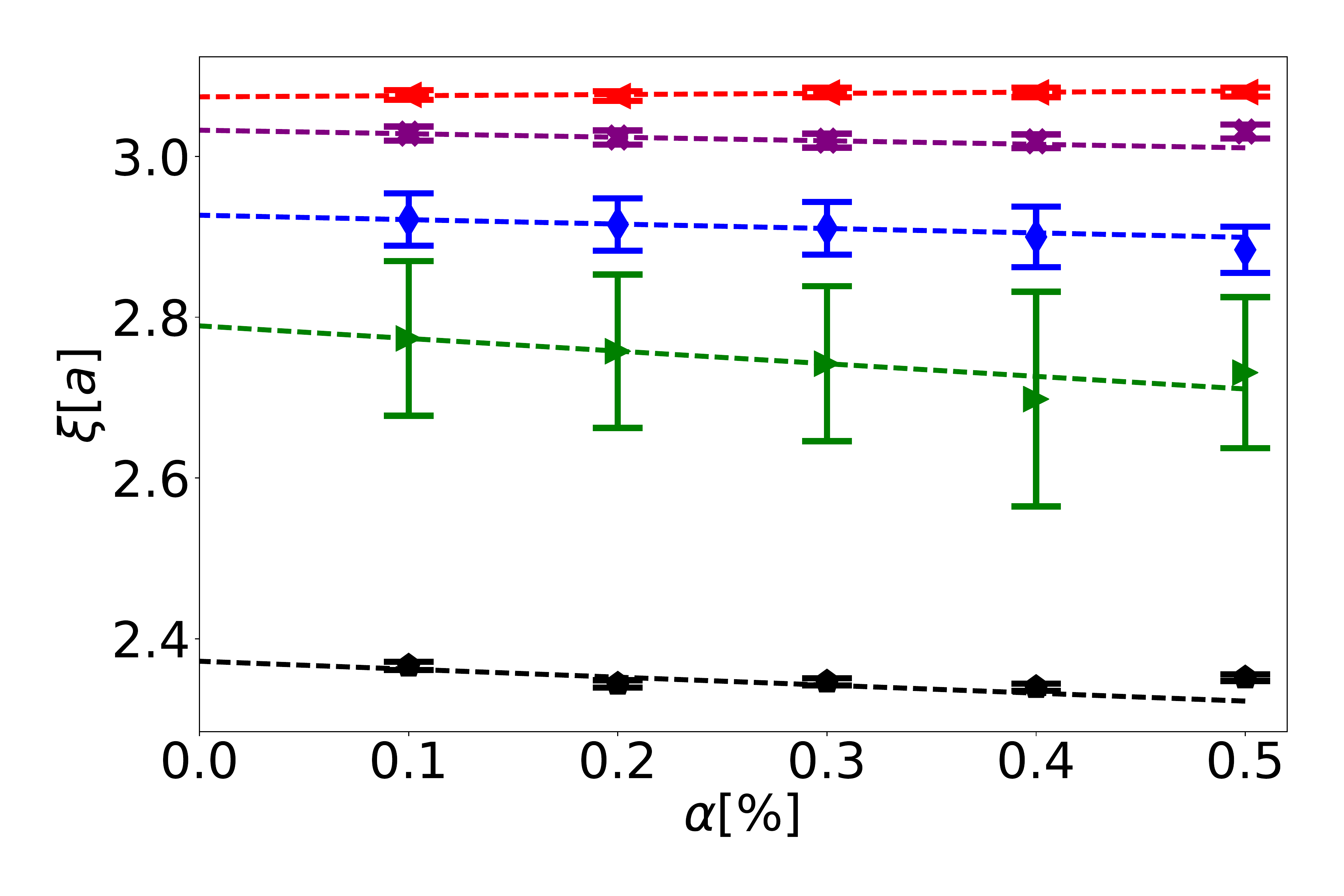}
\caption{
{\color{black}
Development of $\xi$ with cutoff $\alpha$ for disorder strengths $W=0.5$ (green), $1.5$ (blue), $2.0$ (purple), $2.5$(red), $3.5$ (black); error bars depict the uncertainty stemming from the ensemble average.
The dashed lines show a linear fit accounting for the three smallest $\alpha$ values. 
(Parameters: $U{=}1.5$ $L{=}192$; $\NE\approx600-1000$)}
\label{f11}
}
\end{figure}

{\color{black} In Fig. \ref{f11} the dependence of $\xi$ on $\alpha$ at fixed $W$ is shown. 
The data demonstrates good convergence behavior of $\xi$ in terms of the cutoff-parameter $\alpha$; 
in particular, the $\alpha$-dependency of $\xi$ is seen to be small as compared to the variation with $W$. 
Figure \ref{f12} re-plots the data shown in Fig. \ref{f11}, so the evolution of $\xi$ with $W$ 
is more clearly illustrated. In particular, it is seen that the non-monotonic behavior is very well 
converged in the cutoff $\alpha$. The stronger change of $\xi$ with $\alpha$ seen at low disorder strengths, 
e.g. at $W=0.05,0.5$, is related to the fact that the distribution of local values, 
$\Delta(\br)$ is narrow at small $W$. In this case, the convergence requirement 
allowing for a maximal percentage  $\alpha$ of change from cycle to cycle has implications 
for a substantial fraction of all sites; with broad distributions, convergence of most sites will be 
much better than $\alpha$.
\begin{figure}[t]
\includegraphics[width=1.0\linewidth]{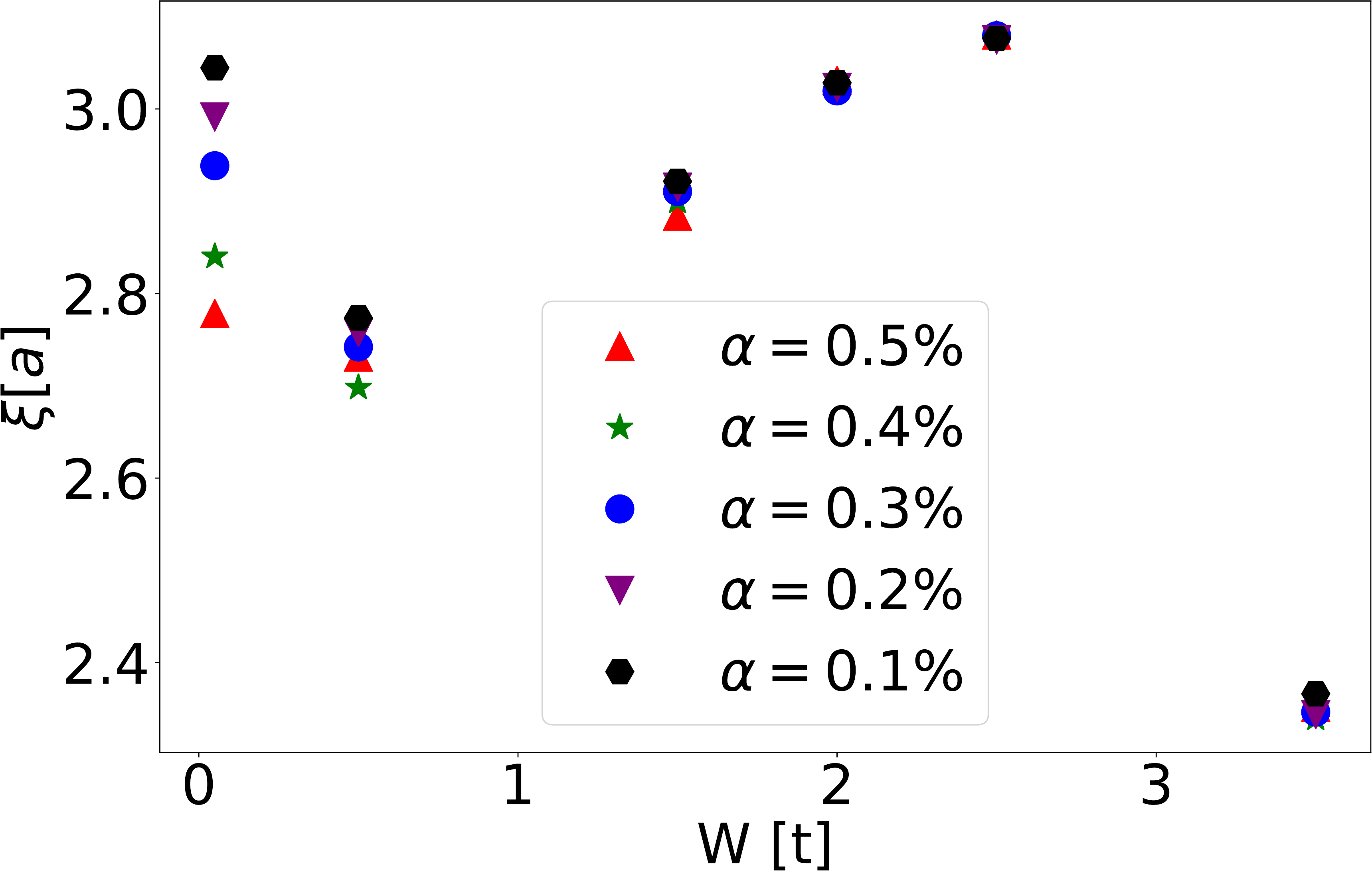}
\caption{
{\color{black}
Re-plot of the data Fig. \eqref{f11} to illustrate the (converged) variation of $\xi$ with $W$. }
\label{f12}
}
\end{figure}

\section*{Acknowledgement}
We are grateful to Soumya Bera, Igor Burmistrov, Christoph Strunk and Thomas Vojta for numerous inspiring discussions; 
we also express our gratitude to Ivan Kondov for sharing mathematical and computational expertise. Support from the DFG under EV30/11-1 and EV30/12-1 
is  acknowledged. The authors gratefully acknowledge the Gauss Centre for Supercomputing e.V. (www.gauss-centre.eu) for funding this project by providing computing time on the GCS Supercomputer SuperMUC at Leibniz Supercomputing Centre (www.lrz.de). This work was performed on the supercomputer ForHLR funded by the Ministry of Science, Research and the Arts Baden-W\"urttemberg and by the Federal Ministry of Education and Research.

%


\begin{thebibliography}{53}%
\makeatletter
\providecommand \@ifxundefined [1]{%
 \@ifx{#1\undefined}
}%
\providecommand \@ifnum [1]{%
 \ifnum #1\expandafter \@firstoftwo
 \else \expandafter \@secondoftwo
 \fi
}%
\providecommand \@ifx [1]{%
 \ifx #1\expandafter \@firstoftwo
 \else \expandafter \@secondoftwo
 \fi
}%
\providecommand \natexlab [1]{#1}%
\providecommand \enquote  [1]{``#1''}%
\providecommand \bibnamefont  [1]{#1}%
\providecommand \bibfnamefont [1]{#1}%
\providecommand \citenamefont [1]{#1}%
\providecommand \href@noop [0]{\@secondoftwo}%
\providecommand \href [0]{\begingroup \@sanitize@url \@href}%
\providecommand \@href[1]{\@@startlink{#1}\@@href}%
\providecommand \@@href[1]{\endgroup#1\@@endlink}%
\providecommand \@sanitize@url [0]{\catcode `\\12\catcode `\$12\catcode
  `\&12\catcode `\#12\catcode `\^12\catcode `\_12\catcode `\%12\relax}%
\providecommand \@@startlink[1]{}%
\providecommand \@@endlink[0]{}%
\providecommand \url  [0]{\begingroup\@sanitize@url \@url }%
\providecommand \@url [1]{\endgroup\@href {#1}{\urlprefix }}%
\providecommand \urlprefix  [0]{URL }%
\providecommand \Eprint [0]{\href }%
\providecommand \doibase [0]{http://dx.doi.org/}%
\providecommand \selectlanguage [0]{\@gobble}%
\providecommand \bibinfo  [0]{\@secondoftwo}%
\providecommand \bibfield  [0]{\@secondoftwo}%
\providecommand \translation [1]{[#1]}%
\providecommand \BibitemOpen [0]{}%
\providecommand \bibitemStop [0]{}%
\providecommand \bibitemNoStop [0]{.\EOS\space}%
\providecommand \EOS [0]{\spacefactor3000\relax}%
\providecommand \BibitemShut  [1]{\csname bibitem#1\endcsname}%
\let\auto@bib@innerbib\@empty
\bibitem [{\citenamefont {Zirnbauer}(1996)}]{Zirnbauer1996ISB}%
  \BibitemOpen
  \bibfield  {author} {\bibinfo {author} {\bibfnamefont {M.~R.}\ \bibnamefont
  {Zirnbauer}},\ }\href@noop {} {\bibfield  {journal} {\bibinfo  {journal} {J.
  Math. Phys.}\ }\textbf {\bibinfo {volume} {37}},\ \bibinfo {pages} {4986}
  (\bibinfo {year} {1996})}\BibitemShut {NoStop}%
\bibitem [{\citenamefont {Altland}\ and\ \citenamefont
  {Zirnbauer}(1997)}]{Zirnbauer1997ISB}%
  \BibitemOpen
  \bibfield  {author} {\bibinfo {author} {\bibfnamefont {A.}~\bibnamefont
  {Altland}}\ and\ \bibinfo {author} {\bibfnamefont {M.~R.}\ \bibnamefont
  {Zirnbauer}},\ }\href@noop {} {\bibfield  {journal} {\bibinfo  {journal}
  {Phys. Rev. B}\ }\textbf {\bibinfo {volume} {55}},\ \bibinfo {pages} {1142}
  (\bibinfo {year} {1997})}\BibitemShut {NoStop}%
\bibitem [{\citenamefont {Heinzner}\ \emph {et~al.}(2005)\citenamefont
  {Heinzner}, \citenamefont {Huckleberry},\ and\ \citenamefont
  {Zirnbauer}}]{Heinzner2005ISB}%
  \BibitemOpen
  \bibfield  {author} {\bibinfo {author} {\bibfnamefont {P.}~\bibnamefont
  {Heinzner}}, \bibinfo {author} {\bibfnamefont {A.}~\bibnamefont
  {Huckleberry}}, \ and\ \bibinfo {author} {\bibfnamefont {M.~R.}\ \bibnamefont
  {Zirnbauer}},\ }\href@noop {} {\bibfield  {journal} {\bibinfo  {journal}
  {Commun. Math. Phys.}\ }\textbf {\bibinfo {volume} {257}},\ \bibinfo {pages}
  {725} (\bibinfo {year} {2005})}\BibitemShut {NoStop}%
\bibitem [{\citenamefont {Mirlin}\ \emph {et~al.}(1996)\citenamefont {Mirlin},
  \citenamefont {Fyodorov}, \citenamefont {Dittes}, \citenamefont {Quezada},\
  and\ \citenamefont {Seligman}}]{Mirlin1996}%
  \BibitemOpen
  \bibfield  {author} {\bibinfo {author} {\bibfnamefont {A.~D.}\ \bibnamefont
  {Mirlin}}, \bibinfo {author} {\bibfnamefont {Y.~V.}\ \bibnamefont
  {Fyodorov}}, \bibinfo {author} {\bibfnamefont {F.-M.}\ \bibnamefont
  {Dittes}}, \bibinfo {author} {\bibfnamefont {J.}~\bibnamefont {Quezada}}, \
  and\ \bibinfo {author} {\bibfnamefont {T.~H.}\ \bibnamefont {Seligman}},\
  }\href@noop {} {\bibfield  {journal} {\bibinfo  {journal} {Phys. Rev. E}\
  }\textbf {\bibinfo {volume} {54}},\ \bibinfo {pages} {3221} (\bibinfo {year}
  {1996})}\BibitemShut {NoStop}%
\bibitem [{\citenamefont {Mirlin}\ and\ \citenamefont
  {Evers}(2000)}]{Mirlin2000b}%
  \BibitemOpen
  \bibfield  {author} {\bibinfo {author} {\bibfnamefont {A.~D.}\ \bibnamefont
  {Mirlin}}\ and\ \bibinfo {author} {\bibfnamefont {F.}~\bibnamefont {Evers}},\
  }\href@noop {} {\bibfield  {journal} {\bibinfo  {journal} {Phys. Rev. B}\
  }\textbf {\bibinfo {volume} {62}},\ \bibinfo {pages} {7920} (\bibinfo {year}
  {2000})}\BibitemShut {NoStop}%
\bibitem [{\citenamefont {Hedin}(1965)}]{hedin1965}%
  \BibitemOpen
  \bibfield  {author} {\bibinfo {author} {\bibfnamefont {L.}~\bibnamefont
  {Hedin}},\ }\href@noop {} {\bibfield  {journal} {\bibinfo  {journal} {Phys.
  Rev.}\ }\textbf {\bibinfo {volume} {139}},\ \bibinfo {pages} {A796} (\bibinfo
  {year} {1965})}\BibitemShut {NoStop}%
\bibitem [{\citenamefont {Bechstedt}(2015)}]{bechstedt}%
  \BibitemOpen
  \bibfield  {author} {\bibinfo {author} {\bibfnamefont {F.}~\bibnamefont
  {Bechstedt}},\ }\href@noop {} {\emph {\bibinfo {title} {Many-Body Approach to
  Electronic Excitations}}}\ (\bibinfo  {publisher} {Berlin: Springer},\
  \bibinfo {year} {2015})\BibitemShut {NoStop}%
\bibitem [{\citenamefont {van Setten}\ \emph {et~al.}(2013)\citenamefont {van
  Setten}, \citenamefont {Weigend},\ and\ \citenamefont
  {Evers}}]{vanSetten2013}%
  \BibitemOpen
  \bibfield  {author} {\bibinfo {author} {\bibfnamefont {M.~J.}\ \bibnamefont
  {van Setten}}, \bibinfo {author} {\bibfnamefont {F.}~\bibnamefont {Weigend}},
  \ and\ \bibinfo {author} {\bibfnamefont {F.}~\bibnamefont {Evers}},\
  }\href@noop {} {\bibfield  {journal} {\bibinfo  {journal} {J. Chem. Theory
  Comput.}\ }\textbf {\bibinfo {volume} {9}},\ \bibinfo {pages} {232} (\bibinfo
  {year} {2013})}\BibitemShut {NoStop}%
\bibitem [{\citenamefont {Yang}\ \emph {et~al.}(1995)\citenamefont {Yang},
  \citenamefont {MacDonald},\ and\ \citenamefont {Huckestein}}]{yang95}%
  \BibitemOpen
  \bibfield  {author} {\bibinfo {author} {\bibfnamefont {E.}~\bibnamefont
  {Yang}}, \bibinfo {author} {\bibfnamefont {A.}~\bibnamefont {MacDonald}}, \
  and\ \bibinfo {author} {\bibfnamefont {B.}~\bibnamefont {Huckestein}},\
  }\href@noop {} {\bibfield  {journal} {\bibinfo  {journal} {Phys. Rev. Lett.}\
  }\textbf {\bibinfo {volume} {74}},\ \bibinfo {pages} {3229} (\bibinfo {year}
  {1995})}\BibitemShut {NoStop}%
\bibitem [{\citenamefont {Huckestein}\ and\ \citenamefont
  {Backhaus}(1999)}]{backhaus99}%
  \BibitemOpen
  \bibfield  {author} {\bibinfo {author} {\bibfnamefont {B.}~\bibnamefont
  {Huckestein}}\ and\ \bibinfo {author} {\bibfnamefont {M.}~\bibnamefont
  {Backhaus}},\ }\href@noop {} {\bibfield  {journal} {\bibinfo  {journal}
  {Phys. Rev. Lett.}\ }\textbf {\bibinfo {volume} {82}},\ \bibinfo {pages}
  {5100} (\bibinfo {year} {1999})}\BibitemShut {NoStop}%
\bibitem [{\citenamefont {Epperlein}\ \emph {et~al.}(1997)\citenamefont
  {Epperlein}, \citenamefont {Schreiber},\ and\ \citenamefont
  {Vojta}}]{epperlein97}%
  \BibitemOpen
  \bibfield  {author} {\bibinfo {author} {\bibfnamefont {F.}~\bibnamefont
  {Epperlein}}, \bibinfo {author} {\bibfnamefont {M.}~\bibnamefont
  {Schreiber}}, \ and\ \bibinfo {author} {\bibfnamefont {T.}~\bibnamefont
  {Vojta}},\ }\href@noop {} {\bibfield  {journal} {\bibinfo  {journal} {Phys.
  Rev. B}\ }\textbf {\bibinfo {volume} {56}},\ \bibinfo {pages} {5890}
  (\bibinfo {year} {1997})}\BibitemShut {NoStop}%
\bibitem [{\citenamefont {Amini}\ \emph {et~al.}(2014)\citenamefont {Amini},
  \citenamefont {Kravtsov},\ and\ \citenamefont {M{\"u}ller}}]{amini14}%
  \BibitemOpen
  \bibfield  {author} {\bibinfo {author} {\bibfnamefont {M.}~\bibnamefont
  {Amini}}, \bibinfo {author} {\bibfnamefont {V.~E.}\ \bibnamefont {Kravtsov}},
  \ and\ \bibinfo {author} {\bibfnamefont {M.}~\bibnamefont {M{\"u}ller}},\
  }\href@noop {} {\bibfield  {journal} {\bibinfo  {journal} {New J. Phys.}\
  }\textbf {\bibinfo {volume} {16}},\ \bibinfo {pages} {015022} (\bibinfo
  {year} {2014})}\BibitemShut {NoStop}%
\bibitem [{\citenamefont {Lee}\ and\ \citenamefont {Kim}(2018)}]{Lee2018}%
  \BibitemOpen
  \bibfield  {author} {\bibinfo {author} {\bibfnamefont {H.-J.}\ \bibnamefont
  {Lee}}\ and\ \bibinfo {author} {\bibfnamefont {K.-S.}\ \bibnamefont {Kim}},\
  }\href@noop {} {\bibfield  {journal} {\bibinfo  {journal} {Phys. Rev. B}\
  }\textbf {\bibinfo {volume} {97}},\ \bibinfo {pages} {155105} (\bibinfo
  {year} {2018})}\BibitemShut {NoStop}%
\bibitem [{\citenamefont {Covaci}\ \emph {et~al.}(2010)\citenamefont {Covaci},
  \citenamefont {Peeters},\ and\ \citenamefont {Berciu}}]{peeters2010}%
  \BibitemOpen
  \bibfield  {author} {\bibinfo {author} {\bibfnamefont {L.}~\bibnamefont
  {Covaci}}, \bibinfo {author} {\bibfnamefont {F.~M.}\ \bibnamefont {Peeters}},
  \ and\ \bibinfo {author} {\bibfnamefont {M.}~\bibnamefont {Berciu}},\
  }\href@noop {} {\bibfield  {journal} {\bibinfo  {journal} {Phys. Rev. Lett.}\
  }\textbf {\bibinfo {volume} {105}},\ \bibinfo {pages} {167006} (\bibinfo
  {year} {2010})}\BibitemShut {NoStop}%
\bibitem [{\citenamefont {Carnio}\ \emph {et~al.}(2019)\citenamefont {Carnio},
  \citenamefont {Hine},\ and\ \citenamefont {R{\"o}mer}}]{carnio2019}%
  \BibitemOpen
  \bibfield  {author} {\bibinfo {author} {\bibfnamefont {E.~G.}\ \bibnamefont
  {Carnio}}, \bibinfo {author} {\bibfnamefont {N.~D.}\ \bibnamefont {Hine}}, \
  and\ \bibinfo {author} {\bibfnamefont {R.~A.}\ \bibnamefont {R{\"o}mer}},\
  }\href@noop {} {\bibfield  {journal} {\bibinfo  {journal} {arXiv:1902.09461}\
  } (\bibinfo {year} {2019})}\BibitemShut {NoStop}%
\bibitem [{\citenamefont {Harashima}\ and\ \citenamefont
  {Slevin}(2012)}]{harashima12}%
  \BibitemOpen
  \bibfield  {author} {\bibinfo {author} {\bibfnamefont {Y.}~\bibnamefont
  {Harashima}}\ and\ \bibinfo {author} {\bibfnamefont {K.}~\bibnamefont
  {Slevin}},\ }\href@noop {} {\bibfield  {journal} {\bibinfo  {journal} {Int.
  J. Mod. Phys. Conf. Ser.}\ }\textbf {\bibinfo {volume} {11}},\ \bibinfo
  {pages} {90} (\bibinfo {year} {2012})}\BibitemShut {NoStop}%
\bibitem [{\citenamefont {Harashima}\ and\ \citenamefont
  {Slevin}(2014)}]{harashima14}%
  \BibitemOpen
  \bibfield  {author} {\bibinfo {author} {\bibfnamefont {Y.}~\bibnamefont
  {Harashima}}\ and\ \bibinfo {author} {\bibfnamefont {K.}~\bibnamefont
  {Slevin}},\ }\href@noop {} {\bibfield  {journal} {\bibinfo  {journal} {Phys.
  Rev. B}\ }\textbf {\bibinfo {volume} {89}},\ \bibinfo {pages} {205108}
  (\bibinfo {year} {2014})}\BibitemShut {NoStop}%
\bibitem [{\citenamefont {Ghosal}\ \emph {et~al.}(1998)\citenamefont {Ghosal},
  \citenamefont {Randeria},\ and\ \citenamefont {Trivedi}}]{ghosal1998}%
  \BibitemOpen
  \bibfield  {author} {\bibinfo {author} {\bibfnamefont {A.}~\bibnamefont
  {Ghosal}}, \bibinfo {author} {\bibfnamefont {M.}~\bibnamefont {Randeria}}, \
  and\ \bibinfo {author} {\bibfnamefont {N.}~\bibnamefont {Trivedi}},\
  }\href@noop {} {\bibfield  {journal} {\bibinfo  {journal} {Phys. Rev. Lett.}\
  }\textbf {\bibinfo {volume} {81}},\ \bibinfo {pages} {3940} (\bibinfo {year}
  {1998})}\BibitemShut {NoStop}%
\bibitem [{\citenamefont {Ghosal}\ \emph {et~al.}(2001)\citenamefont {Ghosal},
  \citenamefont {Randeria},\ and\ \citenamefont {Trivedi}}]{ghosal01}%
  \BibitemOpen
  \bibfield  {author} {\bibinfo {author} {\bibfnamefont {A.}~\bibnamefont
  {Ghosal}}, \bibinfo {author} {\bibfnamefont {M.}~\bibnamefont {Randeria}}, \
  and\ \bibinfo {author} {\bibfnamefont {N.}~\bibnamefont {Trivedi}},\
  }\href@noop {} {\bibfield  {journal} {\bibinfo  {journal} {Phys. Rev. B}\
  }\textbf {\bibinfo {volume} {65}},\ \bibinfo {pages} {014501} (\bibinfo
  {year} {2001})}\BibitemShut {NoStop}%
\bibitem [{\citenamefont {Potirniche}\ \emph {et~al.}(2014)\citenamefont
  {Potirniche}, \citenamefont {Maciejko}, \citenamefont {Nandkishore},\ and\
  \citenamefont {Sondhi}}]{Potirniche2014}%
  \BibitemOpen
  \bibfield  {author} {\bibinfo {author} {\bibfnamefont {I.-D.}\ \bibnamefont
  {Potirniche}}, \bibinfo {author} {\bibfnamefont {J.}~\bibnamefont
  {Maciejko}}, \bibinfo {author} {\bibfnamefont {R.}~\bibnamefont
  {Nandkishore}}, \ and\ \bibinfo {author} {\bibfnamefont {S.~L.}\ \bibnamefont
  {Sondhi}},\ }\href@noop {} {\bibfield  {journal} {\bibinfo  {journal} {Phys.
  Rev. B}\ }\textbf {\bibinfo {volume} {90}},\ \bibinfo {pages} {094516}
  (\bibinfo {year} {2014})}\BibitemShut {NoStop}%
\bibitem [{\citenamefont {Ghosal}\ and\ \citenamefont {Kee}(2004)}]{ghosal04}%
  \BibitemOpen
  \bibfield  {author} {\bibinfo {author} {\bibfnamefont {A.}~\bibnamefont
  {Ghosal}}\ and\ \bibinfo {author} {\bibfnamefont {H.-Y.}\ \bibnamefont
  {Kee}},\ }\href@noop {} {\bibfield  {journal} {\bibinfo  {journal} {Phys.
  Rev. B}\ }\textbf {\bibinfo {volume} {69}},\ \bibinfo {pages} {224513}
  (\bibinfo {year} {2004})}\BibitemShut {NoStop}%
\bibitem [{\citenamefont {Chakraborty}\ \emph {et~al.}(2017)\citenamefont
  {Chakraborty}, \citenamefont {Sensarma},\ and\ \citenamefont
  {Ghosal}}]{ghosal17}%
  \BibitemOpen
  \bibfield  {author} {\bibinfo {author} {\bibfnamefont {D.}~\bibnamefont
  {Chakraborty}}, \bibinfo {author} {\bibfnamefont {R.}~\bibnamefont
  {Sensarma}}, \ and\ \bibinfo {author} {\bibfnamefont {A.}~\bibnamefont
  {Ghosal}},\ }\href@noop {} {\bibfield  {journal} {\bibinfo  {journal} {Phys.
  Rev. B}\ }\textbf {\bibinfo {volume} {95}},\ \bibinfo {pages} {014516}
  (\bibinfo {year} {2017})}\BibitemShut {NoStop}%
\bibitem [{\citenamefont {Ganguly}\ \emph {et~al.}(2017)\citenamefont
  {Ganguly}, \citenamefont {Roy}, \citenamefont {Banerjee}, \citenamefont
  {Singh}, \citenamefont {Ghosal},\ and\ \citenamefont
  {Raychaudhuri}}]{ghosal17b}%
  \BibitemOpen
  \bibfield  {author} {\bibinfo {author} {\bibfnamefont {R.}~\bibnamefont
  {Ganguly}}, \bibinfo {author} {\bibfnamefont {I.}~\bibnamefont {Roy}},
  \bibinfo {author} {\bibfnamefont {A.}~\bibnamefont {Banerjee}}, \bibinfo
  {author} {\bibfnamefont {H.}~\bibnamefont {Singh}}, \bibinfo {author}
  {\bibfnamefont {A.}~\bibnamefont {Ghosal}}, \ and\ \bibinfo {author}
  {\bibfnamefont {P.}~\bibnamefont {Raychaudhuri}},\ }\href@noop {} {\bibfield
  {journal} {\bibinfo  {journal} {Phys. Rev. B}\ }\textbf {\bibinfo {volume}
  {96}},\ \bibinfo {pages} {054509} (\bibinfo {year} {2017})}\BibitemShut
  {NoStop}%
\bibitem [{\citenamefont {Seibold}\ \emph {et~al.}(2012)\citenamefont
  {Seibold}, \citenamefont {Benfatto}, \citenamefont {Castellani},\ and\
  \citenamefont {Lorenzana}}]{castellani12}%
  \BibitemOpen
  \bibfield  {author} {\bibinfo {author} {\bibfnamefont {G.}~\bibnamefont
  {Seibold}}, \bibinfo {author} {\bibfnamefont {L.}~\bibnamefont {Benfatto}},
  \bibinfo {author} {\bibfnamefont {C.}~\bibnamefont {Castellani}}, \ and\
  \bibinfo {author} {\bibfnamefont {J.}~\bibnamefont {Lorenzana}},\ }\href@noop
  {} {\bibfield  {journal} {\bibinfo  {journal} {Phys. Rev. Lett.}\ }\textbf
  {\bibinfo {volume} {108}},\ \bibinfo {pages} {207004} (\bibinfo {year}
  {2012})}\BibitemShut {NoStop}%
\bibitem [{\citenamefont {Lemari{\'e}}\ \emph {et~al.}(2013)\citenamefont
  {Lemari{\'e}}, \citenamefont {Kamlapure}, \citenamefont {Bucheli},
  \citenamefont {Benfatto}, \citenamefont {Lorenzana}, \citenamefont {Seibold},
  \citenamefont {Ganguli}, \citenamefont {Raychaudhuri},\ and\ \citenamefont
  {Castellani}}]{castellani13}%
  \BibitemOpen
  \bibfield  {author} {\bibinfo {author} {\bibfnamefont {G.}~\bibnamefont
  {Lemari{\'e}}}, \bibinfo {author} {\bibfnamefont {A.}~\bibnamefont
  {Kamlapure}}, \bibinfo {author} {\bibfnamefont {D.}~\bibnamefont {Bucheli}},
  \bibinfo {author} {\bibfnamefont {L.}~\bibnamefont {Benfatto}}, \bibinfo
  {author} {\bibfnamefont {J.}~\bibnamefont {Lorenzana}}, \bibinfo {author}
  {\bibfnamefont {G.}~\bibnamefont {Seibold}}, \bibinfo {author} {\bibfnamefont
  {S.~C.}\ \bibnamefont {Ganguli}}, \bibinfo {author} {\bibfnamefont
  {P.}~\bibnamefont {Raychaudhuri}}, \ and\ \bibinfo {author} {\bibfnamefont
  {C.}~\bibnamefont {Castellani}},\ }\href@noop {} {\bibfield  {journal}
  {\bibinfo  {journal} {Phys. Rev. B}\ }\textbf {\bibinfo {volume} {87}},\
  \bibinfo {pages} {184509} (\bibinfo {year} {2013})}\BibitemShut {NoStop}%
\bibitem [{\citenamefont {Cea}\ \emph {et~al.}(2014)\citenamefont {Cea},
  \citenamefont {Bucheli}, \citenamefont {Seibold}, \citenamefont {Benfatto},
  \citenamefont {Lorenzana},\ and\ \citenamefont {Castellani}}]{castellani14}%
  \BibitemOpen
  \bibfield  {author} {\bibinfo {author} {\bibfnamefont {T.}~\bibnamefont
  {Cea}}, \bibinfo {author} {\bibfnamefont {D.}~\bibnamefont {Bucheli}},
  \bibinfo {author} {\bibfnamefont {G.}~\bibnamefont {Seibold}}, \bibinfo
  {author} {\bibfnamefont {L.}~\bibnamefont {Benfatto}}, \bibinfo {author}
  {\bibfnamefont {J.}~\bibnamefont {Lorenzana}}, \ and\ \bibinfo {author}
  {\bibfnamefont {C.}~\bibnamefont {Castellani}},\ }\href@noop {} {\bibfield
  {journal} {\bibinfo  {journal} {Phys. Rev. B}\ }\textbf {\bibinfo {volume}
  {89}},\ \bibinfo {pages} {174506} (\bibinfo {year} {2014})}\BibitemShut
  {NoStop}%
\bibitem [{\citenamefont {Seibold}\ \emph {et~al.}(2015)\citenamefont
  {Seibold}, \citenamefont {Benfatto}, \citenamefont {Castellani},\ and\
  \citenamefont {Lorenzana}}]{castellani15}%
  \BibitemOpen
  \bibfield  {author} {\bibinfo {author} {\bibfnamefont {G.}~\bibnamefont
  {Seibold}}, \bibinfo {author} {\bibfnamefont {L.}~\bibnamefont {Benfatto}},
  \bibinfo {author} {\bibfnamefont {C.}~\bibnamefont {Castellani}}, \ and\
  \bibinfo {author} {\bibfnamefont {J.}~\bibnamefont {Lorenzana}},\ }\href@noop
  {} {\bibfield  {journal} {\bibinfo  {journal} {Phys. Rev. B}\ }\textbf
  {\bibinfo {volume} {92}},\ \bibinfo {pages} {064512} (\bibinfo {year}
  {2015})}\BibitemShut {NoStop}%
\bibitem [{\citenamefont {Dubi}\ \emph {et~al.}(2007)\citenamefont {Dubi},
  \citenamefont {Meir},\ and\ \citenamefont {Avishai}}]{dubi07}%
  \BibitemOpen
  \bibfield  {author} {\bibinfo {author} {\bibfnamefont {Y.}~\bibnamefont
  {Dubi}}, \bibinfo {author} {\bibfnamefont {Y.}~\bibnamefont {Meir}}, \ and\
  \bibinfo {author} {\bibfnamefont {Y.}~\bibnamefont {Avishai}},\ }\href@noop
  {} {\bibfield  {journal} {\bibinfo  {journal} {Nature}\ }\textbf {\bibinfo
  {volume} {449}},\ \bibinfo {pages} {876} (\bibinfo {year}
  {2007})}\BibitemShut {NoStop}%
\bibitem [{\citenamefont {Kadanoff}\ and\ \citenamefont
  {Baym}(1962)}]{kadanoff}%
  \BibitemOpen
  \bibfield  {author} {\bibinfo {author} {\bibfnamefont {L.~P.}\ \bibnamefont
  {Kadanoff}}\ and\ \bibinfo {author} {\bibfnamefont {G.}~\bibnamefont
  {Baym}},\ }\href@noop {} {\emph {\bibinfo {title} {Quantum Statistical
  Mechanics}}}\ (\bibinfo  {publisher} {W.A. Benjamin, Inc.},\ \bibinfo {year}
  {1962})\BibitemShut {NoStop}%
\bibitem [{Notei()}]{Notei}%
  \BibitemOpen
  \bibinfo {note} {Once the scf-field was found an update of $h$ has to be
  computed. This computation is efficiently dealt with by employing the fast
  Fourier-transformation (FFT) and therefore not critical. With FFT an
  operation that formally is ${\protect \cal O}(N_\protect \text {bf}^2)$ can
  be downgraded to ${\protect \cal O}(N_\protect \text {bf}\protect \qopname
  \relax o{ln}N_\protect \text {bf})$.}\BibitemShut {Stop}%
\bibitem [{\citenamefont {Wei{\ss}e}\ \emph {et~al.}(2006)\citenamefont
  {Wei{\ss}e}, \citenamefont {Wellein}, \citenamefont {Alvermann},\ and\
  \citenamefont {Fehske}}]{weisse06}%
  \BibitemOpen
  \bibfield  {author} {\bibinfo {author} {\bibfnamefont {A.}~\bibnamefont
  {Wei{\ss}e}}, \bibinfo {author} {\bibfnamefont {G.}~\bibnamefont {Wellein}},
  \bibinfo {author} {\bibfnamefont {A.}~\bibnamefont {Alvermann}}, \ and\
  \bibinfo {author} {\bibfnamefont {H.}~\bibnamefont {Fehske}},\ }\href@noop {}
  {\bibfield  {journal} {\bibinfo  {journal} {Rev. Mod. Phys.}\ }\textbf
  {\bibinfo {volume} {78}},\ \bibinfo {pages} {275} (\bibinfo {year}
  {2006})}\BibitemShut {NoStop}%
\bibitem [{\citenamefont {Nagai}\ \emph {et~al.}(2012)\citenamefont {Nagai},
  \citenamefont {Ota},\ and\ \citenamefont {Machida}}]{nagai2012}%
  \BibitemOpen
  \bibfield  {author} {\bibinfo {author} {\bibfnamefont {Y.}~\bibnamefont
  {Nagai}}, \bibinfo {author} {\bibfnamefont {Y.}~\bibnamefont {Ota}}, \ and\
  \bibinfo {author} {\bibfnamefont {M.}~\bibnamefont {Machida}},\ }\href@noop
  {} {\bibfield  {journal} {\bibinfo  {journal} {J. Phys. Soc. Jpn}\ }\textbf
  {\bibinfo {volume} {81}},\ \bibinfo {pages} {024710} (\bibinfo {year}
  {2012})}\BibitemShut {NoStop}%
\bibitem [{\citenamefont {Nagai}\ \emph {et~al.}(2013)\citenamefont {Nagai},
  \citenamefont {Shinohara}, \citenamefont {Futamura}, \citenamefont {Ota},\
  and\ \citenamefont {Sakurai}}]{nagai2013}%
  \BibitemOpen
  \bibfield  {author} {\bibinfo {author} {\bibfnamefont {Y.}~\bibnamefont
  {Nagai}}, \bibinfo {author} {\bibfnamefont {Y.}~\bibnamefont {Shinohara}},
  \bibinfo {author} {\bibfnamefont {Y.}~\bibnamefont {Futamura}}, \bibinfo
  {author} {\bibfnamefont {Y.}~\bibnamefont {Ota}}, \ and\ \bibinfo {author}
  {\bibfnamefont {T.}~\bibnamefont {Sakurai}},\ }\href@noop {} {\bibfield
  {journal} {\bibinfo  {journal} {J. Phys. Soc. Jpn}\ }\textbf {\bibinfo
  {volume} {82}},\ \bibinfo {pages} {094701} (\bibinfo {year}
  {2013})}\BibitemShut {NoStop}%
\bibitem [{\citenamefont {Baturina}\ \emph {et~al.}(2007)\citenamefont
  {Baturina}, \citenamefont {Mironov}, \citenamefont {Vinokur}, \citenamefont
  {Baklanov},\ and\ \citenamefont {Strunk}}]{baturina07}%
  \BibitemOpen
  \bibfield  {author} {\bibinfo {author} {\bibfnamefont {T.~I.}\ \bibnamefont
  {Baturina}}, \bibinfo {author} {\bibfnamefont {A.~Y.}\ \bibnamefont
  {Mironov}}, \bibinfo {author} {\bibfnamefont {V.~M.}\ \bibnamefont
  {Vinokur}}, \bibinfo {author} {\bibfnamefont {M.~R.}\ \bibnamefont
  {Baklanov}}, \ and\ \bibinfo {author} {\bibfnamefont {C.}~\bibnamefont
  {Strunk}},\ }\href@noop {} {\bibfield  {journal} {\bibinfo  {journal} {Phys.
  Rev. Lett.}\ }\textbf {\bibinfo {volume} {99}},\ \bibinfo {pages} {257003}
  (\bibinfo {year} {2007})}\BibitemShut {NoStop}%
\bibitem [{\citenamefont {Sac{\'e}p{\'e}}\ \emph {et~al.}(2008)\citenamefont
  {Sac{\'e}p{\'e}}, \citenamefont {Chapelier}, \citenamefont {Baturina},
  \citenamefont {Vinokur}, \citenamefont {Baklanov},\ and\ \citenamefont
  {Sanquer}}]{sacepe08}%
  \BibitemOpen
  \bibfield  {author} {\bibinfo {author} {\bibfnamefont {B.}~\bibnamefont
  {Sac{\'e}p{\'e}}}, \bibinfo {author} {\bibfnamefont {C.}~\bibnamefont
  {Chapelier}}, \bibinfo {author} {\bibfnamefont {T.~I.}\ \bibnamefont
  {Baturina}}, \bibinfo {author} {\bibfnamefont {V.~M.}\ \bibnamefont
  {Vinokur}}, \bibinfo {author} {\bibfnamefont {M.~R.}\ \bibnamefont
  {Baklanov}}, \ and\ \bibinfo {author} {\bibfnamefont {M.}~\bibnamefont
  {Sanquer}},\ }\href@noop {} {\bibfield  {journal} {\bibinfo  {journal} {Phys.
  Rev. Lett.}\ }\textbf {\bibinfo {volume} {101}},\ \bibinfo {pages} {157006}
  (\bibinfo {year} {2008})}\BibitemShut {NoStop}%
\bibitem [{\citenamefont {Bouadim}\ \emph {et~al.}(2011)\citenamefont
  {Bouadim}, \citenamefont {Loh}, \citenamefont {Randeria},\ and\ \citenamefont
  {Trivedi}}]{trivedi11}%
  \BibitemOpen
  \bibfield  {author} {\bibinfo {author} {\bibfnamefont {K.}~\bibnamefont
  {Bouadim}}, \bibinfo {author} {\bibfnamefont {Y.~L.}\ \bibnamefont {Loh}},
  \bibinfo {author} {\bibfnamefont {M.}~\bibnamefont {Randeria}}, \ and\
  \bibinfo {author} {\bibfnamefont {N.}~\bibnamefont {Trivedi}},\ }\href@noop
  {} {\bibfield  {journal} {\bibinfo  {journal} {Nat. Phys.}\ }\textbf
  {\bibinfo {volume} {7}},\ \bibinfo {pages} {884} (\bibinfo {year}
  {2011})}\BibitemShut {NoStop}%
\bibitem [{\citenamefont {Cea}\ \emph {et~al.}(2015)\citenamefont {Cea},
  \citenamefont {Castellani}, \citenamefont {Seibold},\ and\ \citenamefont
  {Benfatto}}]{castellani15b}%
  \BibitemOpen
  \bibfield  {author} {\bibinfo {author} {\bibfnamefont {T.}~\bibnamefont
  {Cea}}, \bibinfo {author} {\bibfnamefont {C.}~\bibnamefont {Castellani}},
  \bibinfo {author} {\bibfnamefont {G.}~\bibnamefont {Seibold}}, \ and\
  \bibinfo {author} {\bibfnamefont {L.}~\bibnamefont {Benfatto}},\ }\href@noop
  {} {\bibfield  {journal} {\bibinfo  {journal} {Phys. Rev. Lett.}\ }\textbf
  {\bibinfo {volume} {115}},\ \bibinfo {pages} {157002} (\bibinfo {year}
  {2015})}\BibitemShut {NoStop}%
\bibitem [{\citenamefont {Loh}\ \emph {et~al.}(2016)\citenamefont {Loh},
  \citenamefont {Randeria}, \citenamefont {Trivedi}, \citenamefont {Chang},\
  and\ \citenamefont {Scalettar}}]{scalettar16}%
  \BibitemOpen
  \bibfield  {author} {\bibinfo {author} {\bibfnamefont {Y.~L.}\ \bibnamefont
  {Loh}}, \bibinfo {author} {\bibfnamefont {M.}~\bibnamefont {Randeria}},
  \bibinfo {author} {\bibfnamefont {N.}~\bibnamefont {Trivedi}}, \bibinfo
  {author} {\bibfnamefont {C.-C.}\ \bibnamefont {Chang}}, \ and\ \bibinfo
  {author} {\bibfnamefont {R.}~\bibnamefont {Scalettar}},\ }\href@noop {}
  {\bibfield  {journal} {\bibinfo  {journal} {Phys. Rev. X}\ }\textbf {\bibinfo
  {volume} {6}},\ \bibinfo {pages} {021029} (\bibinfo {year}
  {2016})}\BibitemShut {NoStop}%
\bibitem [{\citenamefont {{M. V. Feigel'man}}\ \emph
  {et~al.}(2007)\citenamefont {{M. V. Feigel'man}}, \citenamefont {Ioffe},
  \citenamefont {Kravtsov},\ and\ \citenamefont
  {Yuzbashyan}}]{Feigelman2007ISB}%
  \BibitemOpen
  \bibfield  {author} {\bibinfo {author} {\bibnamefont {{M. V. Feigel'man}}},
  \bibinfo {author} {\bibfnamefont {L.~B.}\ \bibnamefont {Ioffe}}, \bibinfo
  {author} {\bibfnamefont {V.~E.}\ \bibnamefont {Kravtsov}}, \ and\ \bibinfo
  {author} {\bibfnamefont {E.~A.}\ \bibnamefont {Yuzbashyan}},\ }\href@noop {}
  {\bibfield  {journal} {\bibinfo  {journal} {Phys. Rev. Lett.}\ }\textbf
  {\bibinfo {volume} {98}},\ \bibinfo {pages} {027001} (\bibinfo {year}
  {2007})}\BibitemShut {NoStop}%
\bibitem [{\citenamefont {{M. V. Feigel'man}}\ \emph
  {et~al.}(2010)\citenamefont {{M. V. Feigel'man}}, \citenamefont {Kravtsov},
  \citenamefont {Ioffe},\ and\ \citenamefont {Cuevas}}]{feigelman10}%
  \BibitemOpen
  \bibfield  {author} {\bibinfo {author} {\bibnamefont {{M. V. Feigel'man}}},
  \bibinfo {author} {\bibfnamefont {V.}~\bibnamefont {Kravtsov}}, \bibinfo
  {author} {\bibfnamefont {L.}~\bibnamefont {Ioffe}}, \ and\ \bibinfo {author}
  {\bibfnamefont {E.}~\bibnamefont {Cuevas}},\ }\href@noop {} {\bibfield
  {journal} {\bibinfo  {journal} {Ann. Phys-New York}\ }\textbf {\bibinfo
  {volume} {325}},\ \bibinfo {pages} {1390} (\bibinfo {year}
  {2010})}\BibitemShut {NoStop}%
\bibitem [{\citenamefont {Feigel'man}\ and\ \citenamefont
  {Ioffe}(2015)}]{feigelman15}%
  \BibitemOpen
  \bibfield  {author} {\bibinfo {author} {\bibfnamefont {M.~V.}\ \bibnamefont
  {Feigel'man}}\ and\ \bibinfo {author} {\bibfnamefont {L.~B.}\ \bibnamefont
  {Ioffe}},\ }\href@noop {} {\bibfield  {journal} {\bibinfo  {journal} {Phys.
  Rev. B}\ }\textbf {\bibinfo {volume} {92}},\ \bibinfo {pages} {100509(R)}
  (\bibinfo {year} {2015})}\BibitemShut {NoStop}%
\bibitem [{\citenamefont {Burmistrov}\ \emph {et~al.}(2012)\citenamefont
  {Burmistrov}, \citenamefont {Gornyi},\ and\ \citenamefont
  {Mirlin}}]{Burmistrov2012ISB}%
  \BibitemOpen
  \bibfield  {author} {\bibinfo {author} {\bibfnamefont {I.}~\bibnamefont
  {Burmistrov}}, \bibinfo {author} {\bibfnamefont {I.}~\bibnamefont {Gornyi}},
  \ and\ \bibinfo {author} {\bibfnamefont {A.}~\bibnamefont {Mirlin}},\
  }\href@noop {} {\bibfield  {journal} {\bibinfo  {journal} {Phys. Rev. Lett.}\
  }\textbf {\bibinfo {volume} {108}},\ \bibinfo {pages} {017002} (\bibinfo
  {year} {2012})}\BibitemShut {NoStop}%
\bibitem [{\citenamefont {Burmistrov}\ \emph {et~al.}(2013)\citenamefont
  {Burmistrov}, \citenamefont {Gornyi},\ and\ \citenamefont
  {Mirlin}}]{Burmistrov2013}%
  \BibitemOpen
  \bibfield  {author} {\bibinfo {author} {\bibfnamefont {I.}~\bibnamefont
  {Burmistrov}}, \bibinfo {author} {\bibfnamefont {I.}~\bibnamefont {Gornyi}},
  \ and\ \bibinfo {author} {\bibfnamefont {A.}~\bibnamefont {Mirlin}},\
  }\href@noop {} {\bibfield  {journal} {\bibinfo  {journal} {Phys. Rev. Lett.}\
  }\textbf {\bibinfo {volume} {111}},\ \bibinfo {pages} {066601} (\bibinfo
  {year} {2013})}\BibitemShut {NoStop}%
\bibitem [{\citenamefont {Burmistrov}\ \emph {et~al.}(2015)\citenamefont
  {Burmistrov}, \citenamefont {Gornyi},\ and\ \citenamefont
  {Mirlin}}]{Burmistrov2015ISB}%
  \BibitemOpen
  \bibfield  {author} {\bibinfo {author} {\bibfnamefont {I.}~\bibnamefont
  {Burmistrov}}, \bibinfo {author} {\bibfnamefont {I.}~\bibnamefont {Gornyi}},
  \ and\ \bibinfo {author} {\bibfnamefont {A.}~\bibnamefont {Mirlin}},\
  }\href@noop {} {\bibfield  {journal} {\bibinfo  {journal} {Phys. Rev. B}\
  }\textbf {\bibinfo {volume} {92}},\ \bibinfo {pages} {014506} (\bibinfo
  {year} {2015})}\BibitemShut {NoStop}%
\bibitem [{\citenamefont {Sac{\'e}p{\'e}}\ \emph {et~al.}(2011)\citenamefont
  {Sac{\'e}p{\'e}}, \citenamefont {Dubouchet}, \citenamefont {Chapelier},
  \citenamefont {Sanquer}, \citenamefont {Ovadia}, \citenamefont {Shahar},
  \citenamefont {{M. V. Feigel'man}},\ and\ \citenamefont
  {Ioffe}}]{sacepe2011}%
  \BibitemOpen
  \bibfield  {author} {\bibinfo {author} {\bibfnamefont {B.}~\bibnamefont
  {Sac{\'e}p{\'e}}}, \bibinfo {author} {\bibfnamefont {T.}~\bibnamefont
  {Dubouchet}}, \bibinfo {author} {\bibfnamefont {C.}~\bibnamefont
  {Chapelier}}, \bibinfo {author} {\bibfnamefont {M.}~\bibnamefont {Sanquer}},
  \bibinfo {author} {\bibfnamefont {M.}~\bibnamefont {Ovadia}}, \bibinfo
  {author} {\bibfnamefont {D.}~\bibnamefont {Shahar}}, \bibinfo {author}
  {\bibnamefont {{M. V. Feigel'man}}}, \ and\ \bibinfo {author} {\bibfnamefont
  {L.}~\bibnamefont {Ioffe}},\ }\href@noop {} {\bibfield  {journal} {\bibinfo
  {journal} {Nat. Phys.}\ }\textbf {\bibinfo {volume} {7}},\ \bibinfo {pages}
  {239} (\bibinfo {year} {2011})}\BibitemShut {NoStop}%
\bibitem [{\citenamefont {Hubbard}(1963)}]{hubbard63}%
  \BibitemOpen
  \bibfield  {author} {\bibinfo {author} {\bibfnamefont {J.}~\bibnamefont
  {Hubbard}},\ }\href@noop {} {\bibfield  {journal} {\bibinfo  {journal} {Proc.
  R. Soc. London, Ser. A}\ }\textbf {\bibinfo {volume} {276}},\ \bibinfo
  {pages} {238} (\bibinfo {year} {1963})}\BibitemShut {NoStop}%
\bibitem [{Noteii()}]{Noteii}%
  \BibitemOpen
  \bibinfo {note} {The filling factor is chosen in a way to be close to
  half-filling, which favors a high particle-hole overlap, while avoiding the
  ground state degeneracy of the superconducting state with a charge density
  wave state at half-filling.}\BibitemShut {Stop}%
\bibitem [{Noteiii()}]{Noteiii}%
  \BibitemOpen
  \bibinfo {note} {The datatype for values is double and for the indices is
  integer. Note, that the speed-up to be expected from the matrix-free
  implementation is less than a factor of 9. This is because not only the
  matrix but also the basis vectors have to be loaded from memory, so the
  reduction of memory load operations also depends on how many basis vectors
  are acted on in parallel.}\BibitemShut {Stop}%
\bibitem [{\citenamefont {Pieper}\ \emph {et~al.}(2017)\citenamefont {Pieper},
  \citenamefont {Hager},\ and\ \citenamefont {Fehske}}]{pieper17}%
  \BibitemOpen
  \bibfield  {author} {\bibinfo {author} {\bibfnamefont {A.}~\bibnamefont
  {Pieper}}, \bibinfo {author} {\bibfnamefont {G.}~\bibnamefont {Hager}}, \
  and\ \bibinfo {author} {\bibfnamefont {H.}~\bibnamefont {Fehske}},\
  }\href@noop {} {\bibfield  {journal} {\bibinfo  {journal}
  {arXiv:1708.09689v2}\ } (\bibinfo {year} {2017})}\BibitemShut {NoStop}%
\bibitem [{\citenamefont {Mirlin}(2000)}]{Mirlin2000}%
  \BibitemOpen
  \bibfield  {author} {\bibinfo {author} {\bibfnamefont {A.~D.}\ \bibnamefont
  {Mirlin}},\ }\href@noop {} {\bibfield  {journal} {\bibinfo  {journal} {Phys.
  Rep.}\ }\textbf {\bibinfo {volume} {326}},\ \bibinfo {pages} {259} (\bibinfo
  {year} {2000})}\BibitemShut {NoStop}%
\bibitem [{\citenamefont {Burmistrov}\ \emph {et~al.}(2016)\citenamefont
  {Burmistrov}, \citenamefont {Gornyi},\ and\ \citenamefont
  {Mirlin}}]{Burmistrov2016}%
  \BibitemOpen
  \bibfield  {author} {\bibinfo {author} {\bibfnamefont {I.}~\bibnamefont
  {Burmistrov}}, \bibinfo {author} {\bibfnamefont {I.}~\bibnamefont {Gornyi}},
  \ and\ \bibinfo {author} {\bibfnamefont {A.}~\bibnamefont {Mirlin}},\
  }\href@noop {} {\bibfield  {journal} {\bibinfo  {journal} {Phys. Rev. B}\
  }\textbf {\bibinfo {volume} {93}},\ \bibinfo {pages} {205432} (\bibinfo
  {year} {2016})}\BibitemShut {NoStop}%
\bibitem [{\citenamefont {Stosiek}\ \emph {et~al.}()\citenamefont {Stosiek},
  \citenamefont {Burmistrov},\ and\ \citenamefont
  {Evers}}]{stosiekUnpublished}%
  \BibitemOpen
  \bibfield  {author} {\bibinfo {author} {\bibfnamefont {M.}~\bibnamefont
  {Stosiek}}, \bibinfo {author} {\bibfnamefont {I.}~\bibnamefont {Burmistrov}},
  \ and\ \bibinfo {author} {\bibfnamefont {F.}~\bibnamefont {Evers}},\
  }\href@noop {} {\bibinfo  {journal} {unpublished}\ }\BibitemShut {NoStop}%
\bibitem [{\citenamefont {B.~Bulka}(1987)}]{kramer87}%
  \BibitemOpen
\bibfield  {journal} {  }\bibfield  {author} {\bibinfo {author} {\bibfnamefont
  {B.~K.}\ \bibnamefont {B.~Bulka}, \bibfnamefont {M.~Schreiber}},\ }\href@noop
  {} {\bibfield  {journal} {\bibinfo  {journal} {B.Z. Physik B - Condensed
  Matter}\ }\textbf {\bibinfo {volume} {66}},\ \bibinfo {pages} {21} (\bibinfo
  {year} {1987})}\BibitemShut {NoStop}%
\end{thebibliography}
\end{document}